\documentclass{agujournal2019}
\usepackage{url} %
\usepackage{amsmath}
\usepackage[inline]{trackchanges} %
\usepackage{soul}
\usepackage{natbib}
\graphicspath{{./images/}}
\usepackage[colorinlistoftodos]{todonotes}

\usepackage{hyperref}
\hypersetup{colorlinks,linkcolor={blue!30!black},citecolor={blue!30!black},urlcolor={magenta}}

\draftfalse

\journalname{Journal of Advances in Modeling Earth Systems (JAMES)}

\begin{document}

\title{Parameterizing Vertical Mixing Coefficients in the Ocean Surface Boundary Layer using Neural Networks}

\authors{Aakash Sane \affil{1}, Brandon G. Reichl \affil{2}, Alistair Adcroft \affil{1}, Laure Zanna \affil{3} }

\affiliation{1}{Atmospheric and Oceanic Sciences Program, Princeton University}
\affiliation{2}{NOAA - Geophysical Fluids Dynamics Laboratory}
\affiliation{3}{Courant Institute, New York University}

\correspondingauthor{Aakash Sane}{aakash.sane@princeton.edu}

\begin{keypoints}
\item We improve a parameterization of vertical mixing in the ocean surface boundary layer using neural networks.
\item Neural networks are trained to predict the diffusivity of second moment closure and maintain energetic constraints of the original parameterization. 
\item The improved scheme reduces biases of mixed layer depth and thermocline in an atmospherically forced ocean model. 
\end{keypoints}

\begin{abstract}

Vertical mixing parameterizations in ocean models are formulated on the basis of the physical principles that govern turbulent mixing.  However, many parameterizations include ad hoc components that are not well constrained by theory or data.  One such component is the eddy diffusivity model, where vertical turbulent fluxes of a quantity are parameterized from a variable eddy diffusion coefficient and the mean vertical gradient of the quantity. In this work, we improve a parameterization of vertical mixing in the ocean surface boundary layer by enhancing its eddy diffusivity model using data-driven methods, specifically neural networks.  The neural networks are designed to take extrinsic and intrinsic forcing parameters as input to predict the eddy diffusivity profile and are trained using output data from a second moment closure turbulent mixing scheme. The modified vertical mixing scheme predicts the eddy diffusivity profile through online inference of neural networks and maintains the conservation principles of the standard ocean model equations, which is particularly important for its targeted use in climate simulations. We describe the development and stable implementation of neural networks in an ocean general circulation model and demonstrate that the enhanced scheme outperforms its predecessor by reducing biases in the mixed-layer depth and upper ocean stratification.  Our results demonstrate the potential for data-driven physics-aware parameterizations to improve global climate models.

\end{abstract}

\section*{Plain Language Summary}
The upper region of the ocean is highly energetic and is responsible for transferring mass, energy and biogeochemical traces between the atmosphere and the deeper regions of the ocean. This transport takes place because of turbulent swirling motions, which are found to be of varying sizes. Climate models cannot represent all of these motions because smaller-scale swirls are complex and require additional computational resources. As we cannot neglect those small swirls, we try to approximate their effects on larger-scale motions using mathematical models. These models have a few ad hoc or empirical assumptions that lead to uncertainty when these climate models are used to project the future climate.  To reduce this uncertainty, we augment an existing model of turbulent swirling process with machine learning, which replaces some ad hoc approximations with data-driven neural networks. Neural networks can learn those missing processes more accurately than a traditional physics-based model. The neural networks are shown to improve physics in climate simulations. Although we only touch on one component in an ocean climate model, this approach can be replicated to improve any other component that was using ad hoc assumptions and replace them with data-driven models using techniques from machine learning.

\section{Introduction}

Vertical mixing parameterizations used in ocean general circulation models (OGCMs) represent the effects of unresolved processes on the mean state. These parameterizations have theoretical deficiencies due to the lack of understanding of inadequately represented or missing processes. To overcome this deficiency, parameterizations often require ad hoc/empirical modifications either to approximate the missing processes or to fit data. Vertical mixing schemes can be constructed with various assumptions and different schemes are calibrated differently. These inconsistencies cause the schemes to disagree among themselves \citep{Li2019} and are a major source of model uncertainty \citep{HawkinsSutton,huber2017drivers,Baylor2019,todd2020ocean,Gutjahr2021}. Poorly parameterized mixing can result in errors that accumulate over time, leading to biases in the OGCM. 

New approaches are emerging to improve various parameterizations in ocean and atmosphere models using machine learning. We have applied neural networks, a type of machine learning, to improve a vertical mixing parameterization of the ocean surface boundary layer (OSBL). OSBL is a vital region of turbulence in the ocean. It acts as an interface between the atmosphere and the deeper ocean and it is important to accurately represent mixing in the OSBL. The atmosphere energizes the ocean through the OSBL. Mass, tracers, and momentum are transferred between the atmosphere and deep ocean via the OSBL, and inaccuracies in vertical mixing parameterizations can give rise to uncertain estimates of heat transport, sea level rise, ocean carbon uptake, etc. Including missing processes in upper ocean vertical mixing schemes impact large-scale phenomena, for example, accounting for Langmuir turbulence and submesoscale effects in the OSBL improves simulations of the Indian monsoon \citep{Orenstein2022}. %

\subsection{Modeling vertical diffusivity within Ocean Surface Boundary Layer (OSBL) parameterizations and the assumption of a `universal' shape function}
 
We focus on the energetic Planetary Boundary Layer (ePBL) scheme, a first-order OSBL turbulent mixing parameterization as described in \cite{Reichl2018} (see Section \ref{sec:ePBL}). 
The variation of the vertical diffusivity profile $\kappa_\phi$ (of arbitrary scalar, $\phi$) within the OSBL in ePBL and similar first order schemes can be expressed as a diffusivity scale ($\hat{\kappa}_\phi$) multiplied by a prescribed normalized diffusivity profile (i.e. shape function):
\begin{linenomath*}
\begin{equation}
\kappa_{\phi}\left(\sigma\right) = \hat{\kappa}_\phi g\left(\sigma\right),
\label{eq:Generic_Diffusivity}
\end{equation} 
\end{linenomath*}
where $\hat{\kappa}_\phi$ is often decomposed into a velocity and length scale \citep{Large1994}, $g(\sigma)$ is a dimensionless shape-function, and $\sigma=z/h$ is a dimensionless vertical coordinate, where $z$ is the vertical coordinate and $h$ is the depth of the boundary layer.
OSBL parameterizations that follow this approach traditionally assume that $g(\sigma)$ is a universal function or has a fixed component such as a cubic polynomial that does not change \citep{OBrien1970,Large1994}, and therefore is ad-hoc. In the KPP scheme of \cite{Large1994}, there is a cubic polynomial which is multiplied by a vertically varying turbulent velocity that sets the structure of $\kappa_{\phi}$. The cubic polynomial is universal, whereas turbulent velocity mostly affects the surface layer defined by the region $0<\sigma<0.1$, making the cubic structure dominant below the surface layer. In ePBL scheme \citep{Reichl2018}, $\kappa_{\phi}$ follows similar design.   However, there is no physics-based justification for a universal or ad-hoc profile to exist, and it is widely understood that characteristics of boundary layers can vary considerably with forcing conditions \citep{Li2019}.
We hypothesize that capturing variations of the shape function that are not considered in first-order OSBL schemes such as ePBL will improve the overall representation of vertical mixing in ocean models. In the subsequent text, our usage of the term `universal shape function' will include shape functions which involve some ad-hoc components or approximations such as used in the ePBL scheme (see Section \ref{sec:ePBL}).

\subsection{Second Moment Closure and an alternative to the `universal' shape function}

\begin{figure}[ht!]
    \centering
    \includegraphics[width=0.3\textwidth]{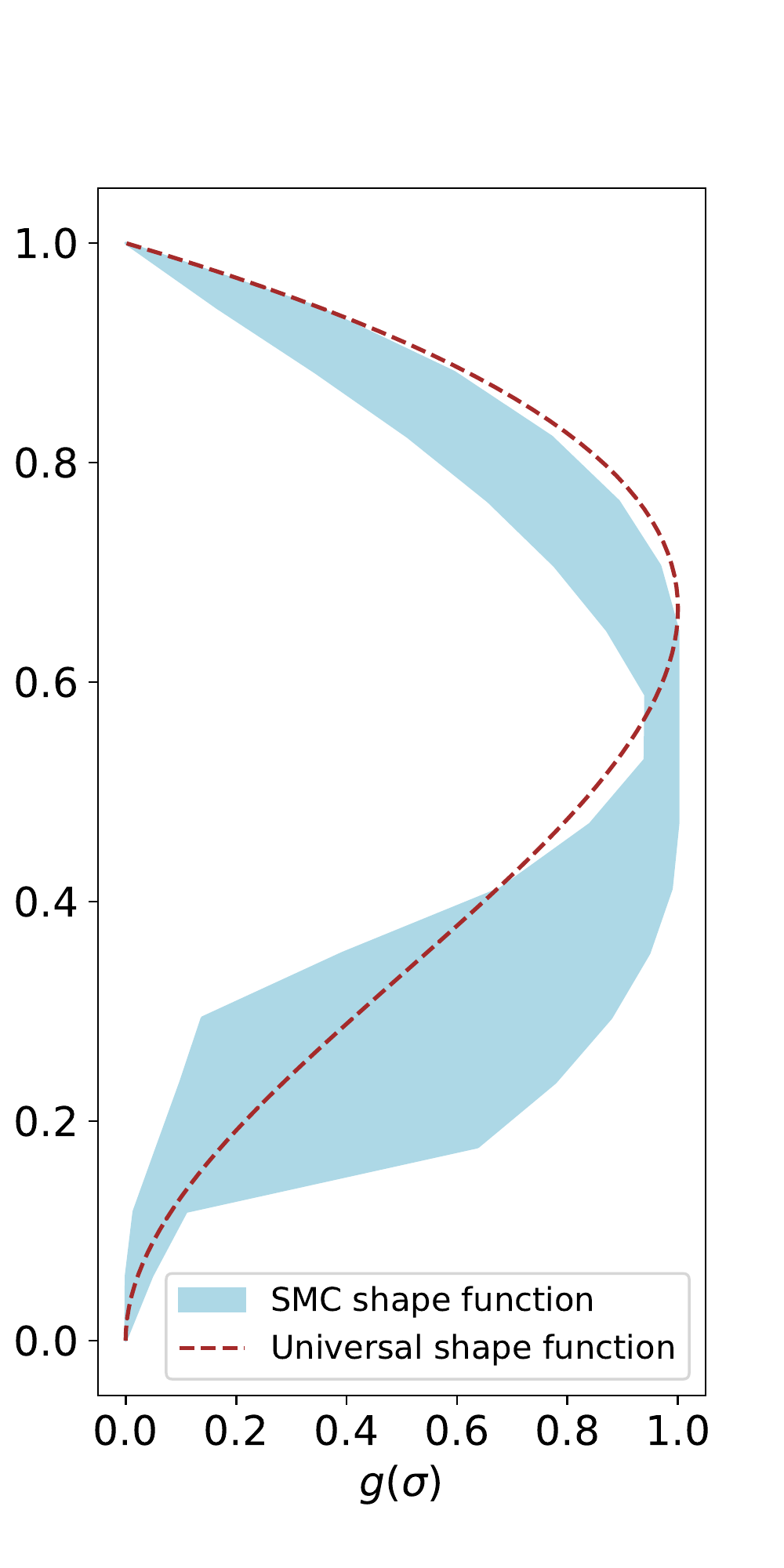}
    \caption{Shape functions derived from various forcing conditions from a second moment closure (blue, shaded region) plotted against a universal shape function (brown, dashed line)  used in GCM vertical mixing schemes. The observed discrepancy between them reveals a limitation in existing vertical mixing schemes. However, this deficiency can be effectively addressed through the application of neural networks, which have the potential to predict the shape function and diffusivity associated with second moment closures.}
    \label{fig:sf_variations}
\end{figure}

Second Moment Closure (SMC) is an alternative approach to predict vertical diffusivity profiles within the OSBL \citep{rodi1987examples,umlauf2005}.
SMC does not require a shape function because it instead predicts the diffusivity from the turbulent kinetic energy ($k$) and the turbulent length scale ($l_t$).
Various SMC approaches exist to predict $k$ and $l_t$ and a general formulation to infer diffusivity is expressed as:
\begin{linenomath*}
\begin{equation}
\kappa_{\phi}\left(z\right) \,=\, c_\phi \, k^{1/2}(z)\, l_t(z),
\label{eq:SMC_Diffusivity}
\end{equation} 
\end{linenomath*}
where $c_\phi$ represents the model stability functions \citep{umlauf2005}.

SMC predicts a profile of vertical diffusivity based on models of physical processes that drive turbulent fluxes within the OSBL.
SMC does not prescribe a shape function {\it a priori}.
However, since SMC directly evaluates a diffusivity profile, the implied shape function and diffusivity scale can be diagnosed from the output. The implied shape function differs significantly from a universal shape function, as seen in Figure \ref{fig:sf_variations}. The diagnosed shape-function and diffusivity scale from SMC can then be used to build a model for use in ePBL. We selected SMC over large eddy simulation as our ``truth" because it is inexpensive compared to the latter, leading to effortless creation of training dataset spanning a wide range of forcing regimes. This is required for machine learning applications as they are hungry for a large amount of data.

A natural question is why SMC is not directly used instead of ePBL in OGCM.
It remains impractical to directly use SMC for vertical mixing in climate simulation due to the sensitivity of their predictions to long time steps and coarse vertical grids often used in climate models \cite[see][]{Reichl2018}.  
However, using the framework described in this article, ePBL can yield a closer approximation to the vertical diffusivity from the SMC scheme without sensitivity to the model's vertical resolution and time step.  Our neural network approach allows ePBL to consider the physics-based variation in the shape function seen in SMC due to solving $k$ and $l_t$. This variability in the shape function will lead to different profiles of vertical mixing within ePBL than using a prescribed universal profile.

\subsection{Machine learning is an emerging tool to improve OGCMs}

Consider a physics-based parameterization that gives an output $\Psi$ as some functional relationship $\mathcal{F}$ between physical quantities $\mathbf{x}$:
\begin{linenomath*}
\begin{equation}
\Psi~=~\mathcal{F}(\mathbf{x}).
\end{equation}
\end{linenomath*}
Finding $\mathcal{F}$ is an optimization problem. It can be set as an optimal linear fit to some combination of $\mathbf{x}$, but the fit might not work for different regimes or might implicitly depend on higher-order combinations of terms in $\mathbf{x}$ (nonlinearity) or some other neglected terms. $\mathcal{F}$ can be assumed to be a function of non-dimensional parameters requiring onerous fitting. With machine learning, $\mathcal{F}$ can be a function of multiple combinations of parameters:
\begin{linenomath*}
\begin{equation}
\mathcal{F}(\mathbf{x}) = \mathcal{N}^{\mathbf{w}} (\mathbf{x}), \end{equation}
\end{linenomath*}
where $\mathcal{N}$ is a machine-learning function, $\mathbf{x}=(x_1, x_2, ...)$ is the input vector and $\mathbf{w}$ are parameters (weights and biases). Machine learning involves determining (learning) the correct values of $\mathbf{w}$ by tuning the hyperparameters that give the optimal $\mathcal{N}$ \citep{Brenner2019}, which is becoming routine due to advances in training algorithms. The machine learning approach provides an avenue to include as many relevant parameters as desired in the vector $\mathbf{x}$, which has been a significant challenge in traditional physics-based approaches.%

Machine learning is favorable for the development and application of climate models due to the abundance  of optimization algorithms and hardware \citep{Balaji2022,christensen2022parametrization}. Studies show that neural networks can be used in idealized model configurations, and recently, the use of machine learning has emerged in realistic GCMs. Artificial neural networks have been shown to improve sub-grid momentum transport in atmospheric models \citep{yuval2021}, predict precipitation \citep{Shamekh2022} and fluxes \citep{Shamekh2023}, while in ocean models they have been used to improve the parameterization of free convection \citep{ramadhan2023}. %
\cite{liang2022exploring} applied deep neural networks to predict temperature and salinity evolution in the OSBL at a weather station (Station Papa). \cite{partee2022} trained a deep neural network to learn subgrid kinetic energy of oceanic mesoscale eddies from a high resolution OGCM to improve their representation in a lower resolution OGCM.
Convolutional neural networks (CNNs) have been used to predict parameterizations of ocean momentum backscatter in a variety of models \citep{bolton2019,zanna2020data, GZ2021} and have been implemented in an ocean primitive equation model \citep{Zhang2023}. \cite{gregory2023} recently employed CNNs to learn data assimilation increments for sea-ice and showed that networks could be used to reduce biases in sea-ice.

Apart from neural networks, techniques considered part of the machine learning toolbox show potential to improve GCMs. The random forest algorithm has been used to parameterize moist convection \citep{Gorman2018} and to learn small-scale processes from a high resolution atmospheric model \citep{yuval2020}. \cite{Mansfield2023} used Gaussian Process emulator to tune gravity wave parameterization in an intermediate complexity atmospheric GCM.   \cite{Souze2020} use a Bayesian technique to fine-tune the non-local flux terms of the KPP parameterization of \cite{Large1994}. 

The aforementioned examples show the potential of enhancing conventional physics-based schemes using machine learning techniques. This article draws inspiration from these demonstrations, recognizing the promise of machine learning in advancing ocean model parameterizations and prompting further investigation in this area.

\subsection{Outline to use Neural Networks and output from SMC to improve ePBL}

Artificial neural networks (ANNs) are trained using output from SMC that directly predicts the profile of vertical diffusivity and do not rely on ad hoc shape functions.
As neural networks are powerful approximators, they can model the variability in the vertical diffusivity profiles of the SMC, but we formulate the ANNs to fit within the simplified framework of the first-order ePBL approach.  
Our procedure has the following advantages:
\begin{enumerate}
\item We use the neural networks to modify the vertical diffusion term within ePBL instead of directly predicting turbulent flux time tendencies (e.g. temperature and salinity), guaranteeing that the scheme conserves physical quantities.
\item The neural networks are introduced in a manner that does not interfere with the potential energy-based mixing constraints of the original ePBL scheme, and therefore ePBL's robust numerical implementation is preserved.
\item The ANNs predict quantities used to compute the diffusivity: the non-dimensional structure (shape function) and a turbulent velocity, which simplifies training, implementation, and interpretability versus directly predicting the diffusivities.%
\item ANNs yield strictly positive values of the vertical diffusivity, an important consideration for numerical stability (see section \ref{sec:342}).
\item Our ANNs are as small as possible to balance accuracy and computational costs, as they will be used in climate timescale OGCM simulations. 
\end{enumerate}

We structure the article as follows. Section \ref{sec:ePBL} describes the ePBL scheme and briefly addresses a calibration/tuning problem. Section \ref{sec:ANN} gives details of the network structure and describes the data used to train networks with estimates of uncertainty. Section \ref{sec:implement} provides details on implementing the enhanced ePBL scheme, hereafter called ePBL\_NN. The new improvements in ePBL\_NN are demonstrated online using free-running single-column model experiments (Section \ref{sec:SCM}), and their impact on biases in an existing ocean-ice climate model is assessed in \ref{sec:JRA-forced}. We conclude with a summary and discussion of the broader implications of this work for applying machine learning to improve parameterizations in ocean climate models.

\section{A Physics-Based Vertical Mixing Framework: The energetic Planetary Boundary Layer (ePBL)}
\label{sec:ePBL}

The ePBL framework, as described by \cite{Reichl2018}, is designed for climate applications of OGCMs and emphasizes robust solutions to changes in model time stepping and vertical resolution.  The scheme is simple enough to implement efficiently within implicit diffusion solvers often used in OGCMs while maintaining important physical constraints on ocean mixing.  The ePBL scheme performs with high skill in idealized models and OGCMs \citep{Reichl_Li_2019,Li2019}, and has been implemented in NOAA - Geophysical Fluid Dynamics Laboratory’s MOM6-based climate models: OM4, CM4, and ESM4 \citep{Adcroft2019,Held2019,Dunne2020}.

ePBL builds on the paradigm of bulk mixed layer models \citep{Kraus1967,Niiler1977}, which constrain the boundary layer depth ($h$) via energetic implications of vertical mixing.  The scheme therefore constrains the mixing based on parameterizing the rate by which turbulent kinetic energy is converted to potential energy within the OSBL:%
\begin{linenomath*}
\begin{equation}
\overbrace{\int_{-h}^{0} \min\left(0,\overline{w' b'}\right)\, dz}^{\rm{integrated~PE~ conversion}} = \mathcal{G}\left(\,\overbrace{f}^{\rm Coriolis},\overbrace{u_*}^{\rm Wind},\overbrace{B_0}^{\rm Buoyancy},\overbrace{h}^{\rm BLD},\overbrace{\int_{-h}^{0} \max\left(0,\overline{w' b'}\right)\,dz}^{\rm Convective~PE~release}\right).
\label{eq:w'b'}
\end{equation}
\end{linenomath*}
Here, $h$ is the (positive) depth of the boundary layer as defined in \cite{Reichl_Li_2019}, $\overline{w^\prime b^\prime}$ is the vertical turbulent buoyancy flux, overbar represents an averaging procedure (e.g., over ensembles), and $\mathcal{G}$ is a relation that depends on the Coriolis parameter $f$, surface friction velocity $u_*$, surface buoyancy flux $B_0=\overline{w^\prime b^\prime}_0$, boundary layer depth $h$, and integrated release of potential energy by convective buoyancy fluxes. \cite{Reichl2018} find $\mathcal{G}$ using simulations from single column models using SMC under a range of forcing scenarios.  Later, this function was enhanced to include Langmuir turbulence using large eddy simulations (LES) \citep{Reichl_Li_2019}.  %

ePBL extends the bulk mixed layer formulation to resolve vertical structure within the OSBL by applying a down-gradient flux profile using the vertical diffusivity given by 
\begin{linenomath*}
\begin{equation}
\overline{w^\prime \phi^\prime}= -\kappa_{\phi}\frac{\partial \overline{\phi}}{\partial z},
\end{equation} 
\end{linenomath*}
where $\kappa_{\phi}$ is the variable diffusivity of a scalar $\phi$. The diffusivity varies with depth and is given in the following form:
\begin{linenomath*}
\begin{equation}
\kappa_{\phi}\left(\sigma\right) = L\left(\sigma\right) v_0\left(\sigma\right),
\label{eq:ePBL_k}
\end{equation} 
\end{linenomath*}
 where $L$ and $v_0$ are length and velocity scales. In the present implementation of ePBL \citep{Reichl2018}, the turbulent Prandtl number is assumed to be one and hence the diffusivity and viscosity are identically modified. Both $L$ and $v_0$ are expressed as functions of position $\sigma$ within the boundary layer.  
The length scale in equation \eqref{eq:ePBL_k} is set as \cite[following][]{OBrien1970, Large1994}:
\begin{linenomath*}
\begin{equation}
 L\left(\sigma \right) = \left( z_o + |z| \right) \left( 1-\sigma \right)^{\gamma}.
 \label{eq:lzgamma}
\end{equation} 
\end{linenomath*}
By assuming a fixed constant for $\gamma$, the expressions given by \eqref{eq:ePBL_k} and \eqref{eq:lzgamma} may be expressed in the same form as \eqref{eq:Generic_Diffusivity}, which reveals the role of the shape function as $g(\sigma)=\sigma(1-\sigma)^\gamma$.
$\gamma$ should not be a fixed constant. Constructing $\gamma$ as a data-driven function is challenging and the form $\sigma(1-\sigma)^\gamma$ does not have a physical basis.  %
The velocity scale $v\left(\sigma\right)$ uses a similar formulation motivated to generally agree with the model $k-\epsilon$ \cite [see Eqs. 43-45 in] [] {Reichl2018}.  

Although the integrated mixing in ePBL is constrained via the function $\mathcal{G}$, the stratification resulting from the mixing is sensitive to the assumptions for $\gamma$ and $v_0$ that set the diffusivity profile within the boundary layer.
Differences in the diffusivity profile mean that even when the energetic constraints are accurate, inconsistent OSBL evolution and stratification can emerge when comparing ePBL with SMC such as $k-\epsilon$ (see Figures \ref{fig:surfheat_hov}, \ref{fig:surfheat_hov_lat1}).  In this article, we enhance the physics-based ePBL approach by improving these velocity- and length-scale formulations with artificial neural networks (ANNs, see Section \ref{sec:ANN}).

\section{Artificial Neural Networks and Training Procedures}
\label{sec:ANN}

Artificial Neural Networks (ANNs) are one of the most widely used forms of machine learning models. ANNs are universal approximators and can find hidden nonlinear relations between quantities \citep{Cybenko1989, Hornik1989, Hornik1991}. In this section, we describe the fundamentals of ANNs and provide details describing the training procedures for the neural networks used to supplement ePBL's eddy diffusivity model.

\subsection{Fundamentals}
ANNs consist of nodes arranged in layers. Nodes are elements of a vector $ \textbf{x} $ that constitute a layer. See Figure \ref{fig:neuralnetworks} for a schematic. Each vector is connected to its adjacent vector via a transformation that involves multiplying with coefficients, called weights $ \textbf{w} $, and adding an offset, called biases $ \textbf{b} $. After transforming the vector with weights and biases, a nonlinear operation yields the next vector (or layer). The nonlinear operator is an activation function $\mathcal{A}$. For an input layer consisting of a vector $\mathbf{x}_1$, one hidden layer $\mathbf{x}_2$ and an output layer $\mathbf{y}$, ANN can be written as
\begin{linenomath*}
\begin{equation}
\begin{aligned}
\mathbf{x}_2 &=  \mathcal{A} \left( \mathbf{w}_1 \mathbf{x}_1 + b_1 \right), \\
\mathbf{y} &= \left( \mathbf{w}_2 \mathbf{x}_2 + b_2 \right).
\end{aligned}
\label{eq:ann}
\end{equation}
\end{linenomath*}
Deeper networks are expanded versions of the above Equation set \ref{eq:ann} and are obtained by adding additional layers.
ANNs can capture nonlinear relationships within certain tolerances and can interpolate with high accuracy within the range of training data. We employ ANNs to learn the nonlinear relationship between chosen input parameters, described below, and the vertical diffusivity profile predicted using SMC. %

\subsection{Learning Diffusivity using two Neural networks $\mathcal{N}_1$ and $\mathcal{N}_2$}
To train the ANN model to predict the diffusivity profile, we use the sigma coordinate defined as $\sigma=z/h$.  Therefore, for the surface $\sigma=0$, and at $h$, $\sigma=1$. We define the diffusivity in the sigma coordinate in terms of a velocity scale, $v_{0}$, the boundary layer depth, $h$, and the nondimensional shape function, $g$:
\begin{linenomath*}
\begin{equation}
    \kappa_{\phi}\left(\sigma\right) = g(\sigma) \cdot h \cdot v_0,
    \label{eq:knn}
\end{equation} 
\end{linenomath*}
where $g(\sigma)$ is defined to give values between [0,1]. %
We could have introduced vertical structure in few or all of the terms on the right hand side in Equation \ref{eq:knn}. Instead we use only $g(\sigma)$ to provide vertical structure as we found out that it was convenient to train one profile than two or more. 
The benefit of adopting the sigma coordinate is in removing the dependence on the vertical coordinate (e.g., grid spacing in $z$) that varies in different ocean models. This allows us to train and infer (feed-forward) without depending on the model's vertical grid, which makes it practical to implement in an ocean model with an adaptive vertical grid \cite[e.g.][]{Bleck2002}. 

The velocity scale $v_0$ in Equation \eqref{eq:knn} does not vary with $\sigma$. The entire vertical structure of $\kappa_{\phi}$ is captured by $g(\sigma)$ alone.  
This is in contrast to Equation \ref{eq:ePBL_k} where both the length scale and the velocity scale vary in vertical direction and contribute to the vertical structure of $\kappa_{\phi}$. We made this choice to simplify the approach so that only one neural network is needed to capture the vertical structure of $\kappa_{\phi}$.  %

We choose to define the shape function and velocity scale using two separate neural networks:
\begin{linenomath*}
\begin{equation}
\begin{aligned}
    g(\sigma)&=\mathcal{N}_1(f,B_0,u_*,h), \\
    v_0 &=\mathcal{N}_2(f,B_0,u_*),
\end{aligned}
\label{eq:n1n2}
\end{equation} 
\end{linenomath*}
where $\mathcal{N}_1$ and $\mathcal{N}_2$ represent two distinct neural networks that are trained independently.  $\mathcal{N}_1$ requires inputs $f,B_0,u_*,$ and $h$, while $\mathcal{N}_2$ is found to depend on $f,~B_0,{\rm~and~}u_*$.  
We chose this strategy rather than combining the two outputs into one ANN for a couple of reasons. First, it is straightforward to cleanly diagnose $g(\sigma)$ and $v_0$ from the data, as will be explained in Sections \ref{sec:N1_training} and \ref{sec:N2_training}.  Second, we anticipate that having separate networks will make the individual networks easier to interpret, which allows us to better understand physical processes modeled by the network.

Both neural networks $\mathcal{N}_{1,2}$ are trained using the Pytorch package \citep{pytorch}. Rectified Linear Unit (ReLU) \citep{nair2010} has been used as the activation function due to its simplicity and rapid convergence in training.

 \subsection{Data for training}
 \label{sec:data_train}

The SMC data used to train the networks is generated using the single column model framework implemented in the General Ocean Turbulence Model \cite[GOTM][]{umlauf2005, umlauf2014}.  %
GOTM provides numerous SMC options to predict the fluxes of turbulence and the vertical diffusivity. We employ a two-equation model: $k-\epsilon$, with stability function closure following \cite{schumann1995}. The choice of this specific SMC parameterization is made to be consistent with \cite{Reichl2018}. Vertical mixing parameterizations remain an active research topic, and currently used schemes, including SMC, can exhibit biases in different forcing regimes and regions \citep{Peters2007,Li2019, Damerell2020, Sane2021, Tirodkar2022}.  Any biases in the training data are inherited by the neural networks. However, our neural networks can be trained using the output of different mixing schemes, including the improved schemes developed in future research.

The GOTM column model consists of a vertical grid with forcing applied at the surface grid point. It is applicable for flows with horizontal homogeneity, i.e., horizontal fluxes are zero or constant. GOTM simulations are performed by changing the following parameters: latitude (Coriolis), surface wind stress (surface friction velocity), and surface heat flux (surface buoyancy flux). Salinity is kept constant and temperature is the only active tracer, though the results are general for any combination of buoyancy fields and forcing. Our initial analysis indicates that the diffusivity of $k-\epsilon$, $\kappa_{k\epsilon}$, depends on the Coriolis parameter $f$, the surface buoyancy flux $B_0$, the surface friction velocity $u_*$ and the depth of the boundary layer, $h$. We can only specify $f$, $B_0$, and $u_*$ in single column simulations, and $h$ is diagnosed from the time evolution simulated by GOTM.

Each GOTM case runs with a set of constant forcings. The time step is set at 60 s, and the vertical grid spacing is 1 m. The depth of the column is 800 m. The simulation results for the $k-\epsilon$ model are converged at this time step and resolution \citep[see Figure 1 in][]{Reichl2018}. The initial conditions consist of zero horizontal velocity, the surface temperature is set at 20 \textdegree C, and the initial temperature stratification is set at 0.005 \textdegree C/m. Data was saved at hourly intervals. For every $f$, $B_0$, and $u_*$, we included one hundred instantaneous profiles of diffusivity at each hour from day 2 to day 6 in the training dataset.

We found ($f, B_0, u_*, h$) to strongly affect diffusivity compared to the background stratification established by the initial conditions. Stratification acts as a barrier to the deepening of the mixed layer, and therefore it is challenging to obtain deeper layers with strong stratification at the bottom of the mixed layer, and this limits the generation of training data spanning a wide range of $h$. Therefore, we choose a weak initial stratification.  The effects of stratification on diffusivity in $k-\epsilon$ most directly impact the rate of deepening of the boundary layer (which is already captured by the energetic constraint of the ePBL), compared to the shape function itself.

Table \ref{table:table1} shows the range of forcing parameters of the training data. Forcing range is different for $\mathcal{N}_1$ and $\mathcal{N}_2$ because we found that the shape function does not vary significantly outside the range stated in Table \ref{table:table1}. Hence we do not train on data outside that range, and the inputs to the network can be capped inside the mixing scheme in MOM6. For example, if the wind stress is 1.3 N/m$^2$, capping prevents the wind stress from going beyond 1.2 N/m$^2$ as the shape function does not vary significantly beyond 1.2 N/m$^2$. A similar argument can be made about the surface heat flux. The range selected to perform the sweep has been informed using the observed forcings in the JRA atmospheric reanalysis data set \citep{jra55}. The range shown in Table \ref{table:table1} covers most of the forcing space certainty as explained in \ref{sec:appendixhx}. For $h$, maximum variations for $g(\sigma)$ were observed between 20 m and 300 m and beyond 300 m $g(\sigma)$ is found to vary marginally. Randomizing the training data and splitting it into two sets (train and validation) could result in very similar elements from similar experiments being present in both the train and the validation data.  This is undesirable since a fully independent validation dataset is required to monitor overfitting when training a neural network. To prevent this issue, a validation data set is independently generated to be 10\% the size of the training data using a fully independent set of forcing parameters. No single element between ($f, B_0, u_*$) is common between the training data set and the validation data set to strictly ensure the independence between the training and validation sets. %

\begin{table}[ht]
\caption{Range of parameters used to generate training data. We have added additional details about GOTM runs to the Supplementary section.}
\centering
\begin{tabular}{l c c}
\hline
 Inputs  & $\mathcal{N}_1$ & $\mathcal{N}_2$  \\
\hline
Surface Heat Flux   & -600 to 600 W/m$^2$ & -2000 to 2000 W/m$^2$   \\
Wind stress   & 0 to 1.2 N/m$^2$ & 0 to 20 N/m$^2$  \\
Surface Friction velocity & 0 to 0.034 & 0 to 0.034 \\
Latitude & -90$^o$ to 90$^o$ & -90$^o$ to 90$^o$ \\
Boundary layer depth & 20 m to 300 m & -  \\
Reference density & 1027 kg/m$^3$ & 1027 kg/m$^3$ \\
Specific heat capacity & 3985 & 3985 \\
Equation of state & Linear & Linear \\
Stratification at initial conditions & 0.005 \textdegree K / m & 0.005 \textdegree K / m \\
\hline
\end{tabular}
\label{table:table1}
\end{table}

\subsection{Training $\mathcal{N}_1$}\label{sec:N1_training}

\begin{figure}
    \centering
    \includegraphics[width=1\textwidth]{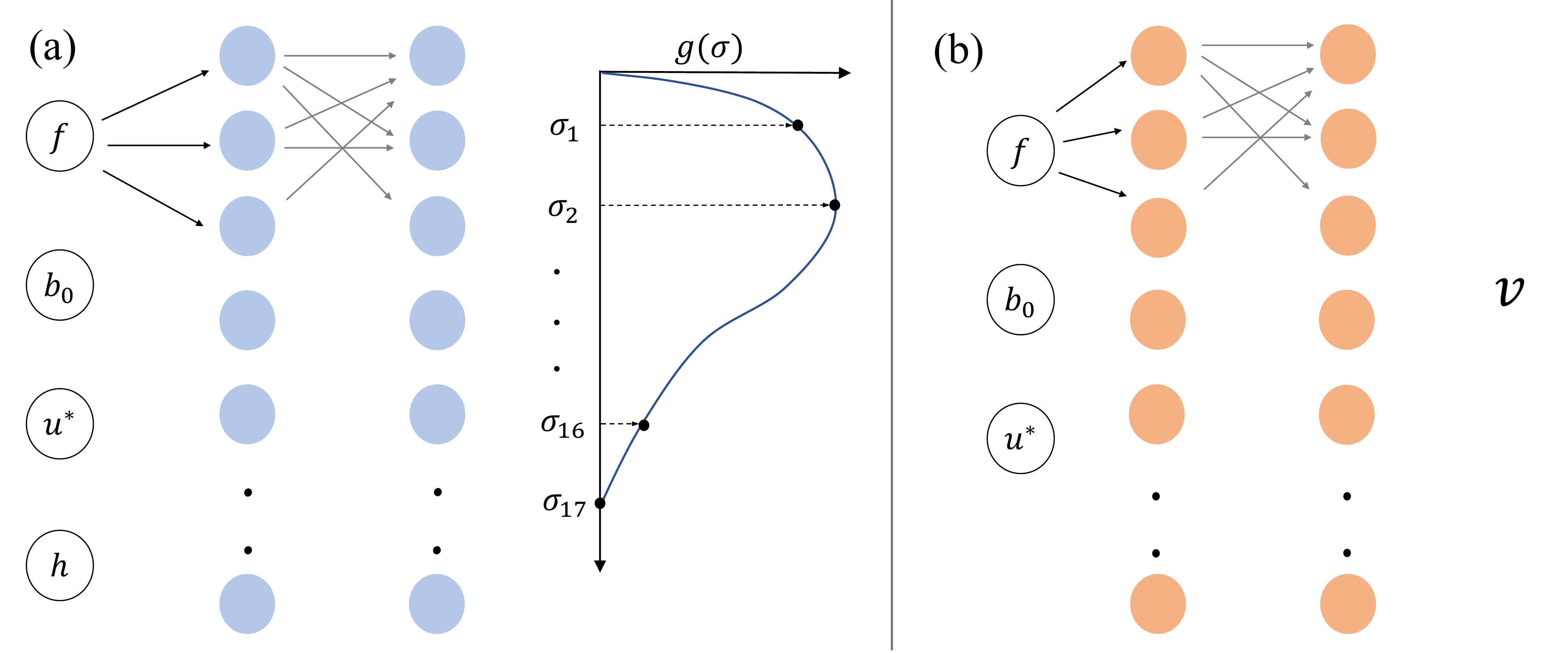}
    \caption{(a) Neural network $\mathcal{N}_1$. It requires four inputs $(f, B_0, u_*, h)$ and output layer consists of 16 nodes giving values of $g(\sigma)$ at those locations. (b) Neural network $\mathcal{N}_2$ requires three inputs $(f, B_0, u_*)$ and output is a scalar velocity scale $v_0$. Diffusivity is obtained by: $\kappa(\sigma)=g(\sigma)\cdot h \cdot v_0$. here, $h$ is the boundary layer depth which is evaluated in the vertical mixing parameterization of OSBL in OGCM using physical arguments.}
    \label{fig:neuralnetworks}
\end{figure}

The parameters $f,B_0,u_*,$ and $h$ are inputs to $\mathcal{N}_1$, while the output consists of a vector having values of $g(\sigma)$ at 16 evenly distributed nodes, as shown in Figure \ref{fig:neuralnetworks}. 
For each set of forcing (i.e. latitude, heat flux, and surface stress), the GOTM output consists of the evolution of the initial conditions into a developed boundary layer. The boundary layer deepens and variations in $\kappa_{k\epsilon}$ emerge. Ignoring the initial $\approx$ 2 days of data, $h$ is diagnosed for each model output with a frequency of 60 minutes by analyzing the profile of the vertical buoyancy flux. Here, $h$ is defined by the depth at which $\overline{w'b'}$ reaches and stays close to zero. This is the maximum extent to which the effect of surface forcing penetrates the upper layer through turbulent buoyancy flux. The diffusivity profile, $\kappa_{k\epsilon}(\sigma)$, is normalized by its maximum to find the shape function:
\begin{linenomath*}
\begin{equation}
g(\sigma)= \kappa_{k\epsilon}(\sigma) / \text{max} (\kappa_{k\epsilon}(\sigma)).
\end{equation}
\end{linenomath*}
The neural network cannot learn a continuous profile in $\sigma$, but instead we train it to learn on a subsampled $g(\sigma)$ grid that consists of 18 equally spaced $\sigma$ points (0, 1/17, 2/17, ...16/17, 1). $\sigma$ at 0 and 1 is ignored in the training because $g(\sigma)=0 $ at the surface ($\sigma=0$). At $\sigma=1 $, $g(\sigma) \rightarrow 0$ and hence is assumed to be zero for training purposes. Therefore, the network predicts $g(\sigma$) at the 16 interior locations. Subsampling $g(\sigma)$ to 18 evenly distributed points was found to be sufficient to capture essential features of $g(\sigma)$ while maintaining a small enough network to later implement in an OGCM. 

\subsubsection{Overcoming limitation in $h$ using synthetic data}

ANNs show high prediction skill when input is within the range of training data.  GOTM experiments can cover a wide range of data points that span latitude, heat flux, and surface wind stress, such as those historically observed in the real ocean. However, this is not true for $h$ as it evolves prognostically and we cannot set its range for each run. We have chosen $h$ to vary from 20 to 300 m (see Table \ref{table:table1}) for training purposes, but for some surface heating conditions the boundary layer depth will saturate towards the Monin-Obukhov length $L_{MO}$, which might be less than 300 m. As $h$ is an input to the network and if for a particular case $L_{MO} < h < 300 $ m, then profiles will not exist and the network will have to predict outside the range of the training dataset. The network might end up predicting spurious profiles. 

We address this issue by supplementing the training with synthetic data. For a particular case, if $h$ saturates to, for example, 200 m, then an additional 10 profiles are added to cover the missing range of 200 m to 300 m in the training data. The shape function for these synthetic profiles is assumed to be the same as when $h=$ 200 m, that is, $g(\sigma)$ for $h \in (200, 300)$ will have the same values as $g(\sigma)$ for $h=200$~m. This assumption is reasonable, since $g(\sigma)$ was found to vary little for deeper boundary layers with surface heating. 

Strong convection can cause a related issue due to quick deepening of the boundary layer within the spin-up phase of the turbulent OSBL. This gap is filled in the same way as described for deep boundary layer gaps. If the lowest value of $h$ is, for example, 100 m, then ten profiles are added that cover 20 m to 100 m. The shape function for these ten profiles is assumed to be the same as that when $h=$
 100 m. 
This fill-up of gaps in $h$ is necessary to stabilize ANNs trained with our existing datasets. Knowing the exact bounding box of the training data set is imperative for a successful and stable implementation in a GCM. %

\subsubsection{Forcing $\mathcal{N}_1$ to be strictly positive}
\label{sec:342}
The network $\mathcal{N}_1$ consists of 4 input nodes, two hidden layers, and 16 output nodes (sensitivity to network hyperparameters is described in the next subsection). The four input nodes correspond to $\left(f, B_0, u_*, h\right)$. The output nodes predict the shape function as described above. The output of $\mathcal{N}_1$, $g(\sigma)$, is a vector of length 16. If $g(\sigma)$ predicts a negative value of the shape function for any $\sigma$, it would lead to negative diffusivity values. We prevent this by training on the logarithm of $g(\sigma)$. $\mathcal{N}_1$ predicts $log(g(\sigma))$ and, while inferring, the exponential function is used. This ensures that the shape function is strictly positive.

The four inputs to the network ($f, B_0, u_*, h$) are normalized by their respective mean and standard deviation of the training data. For the 16 output nodes, each output was normalized by its own mean and standard deviation. For output node i, $log(g(\sigma_i))$ was transformed into 
\begin{linenomath*}
\begin{equation}
log(g(\sigma_i)) \rightarrow \frac{log(g(\sigma_i)) - \overline{log(g(\sigma_i))}}{\langle{log(g(\sigma_i))}\rangle}
\end{equation} 
\end{linenomath*} 
before training.
The overbar denotes the mean, and the angled brackets denote the standard deviation. 

\subsubsection{Network Skill and hyperparameter sweep}

To train $\mathcal{N}_1$ two hyperparameters need to be tuned. The number of hidden layers and the number of nodes in those layers. For simplicity, we chose the same number of nodes in each hidden layer. A sweep was performed to test the accuracy of different networks. We varied the number of hidden layers from 2 to 4 and the number of hidden nodes in each layer from 2 to 512. Training data was randomized and provided as a single batch to train networks.

\begin{figure}[ht]
    \centering
    \includegraphics[width=1.0\linewidth]{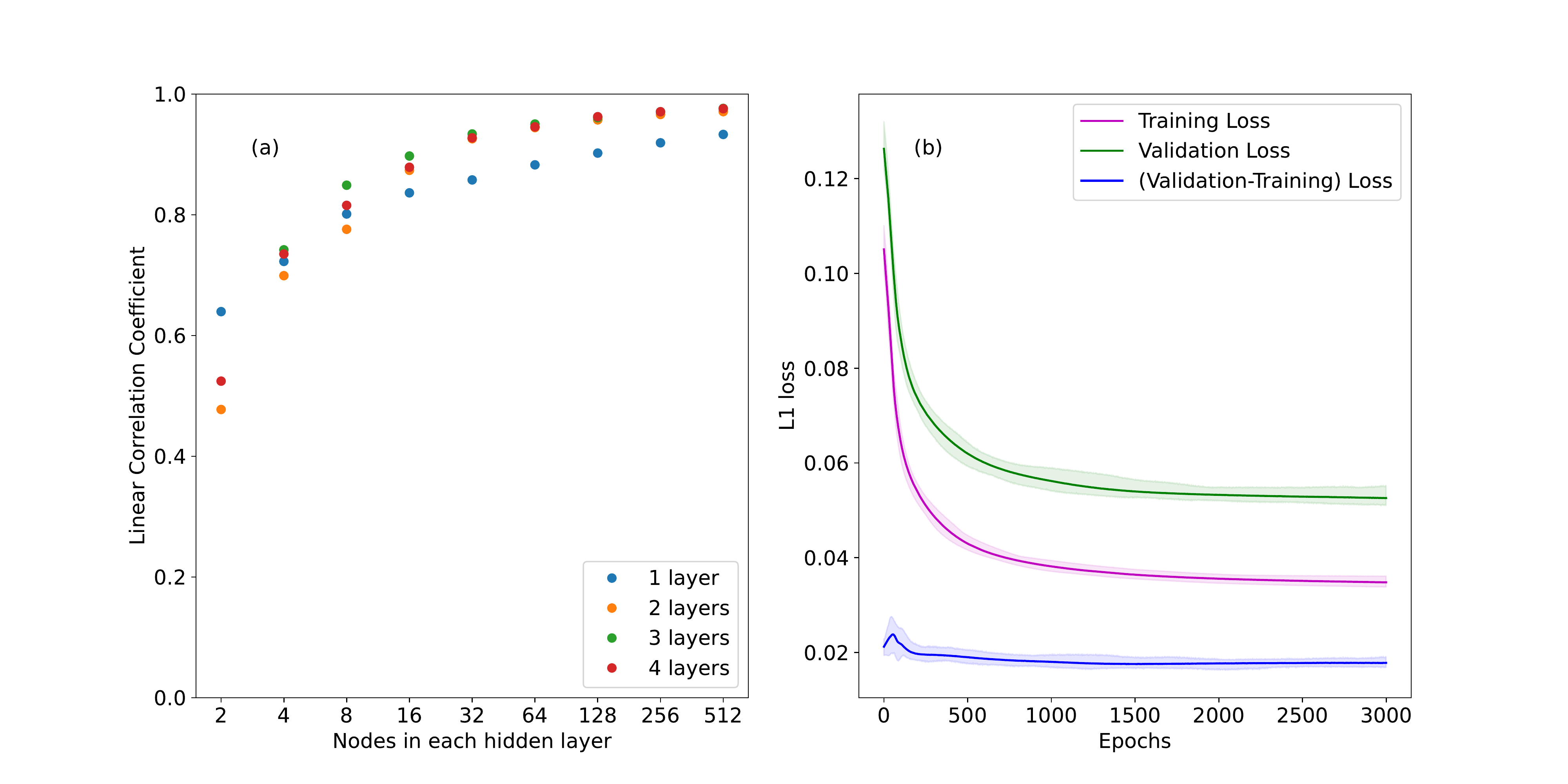}
    \caption{(a) Average linear correlation coefficient between network output and the true values from data. Averaging has been done over all the 16 nodes. Different color represent different number of layers and x-axis shows the nodes in each hidden layer. (b) L1 loss curves for training and validation.}
\label{fig:r2scores}
\end{figure}

To measure the network's performance, linear correlation coefficient between the validation data and its prediction was calculated (see Figure \ref{fig:r2scores} (a)) . The linear correlations for the 16 nodes were weight averaged with the mean value of $g(\sigma)$ of the training data. The weight-averaged correlation is a better estimate of the network's skill for the given set of hidden nodes, as it reduces the influence of noisy values at the bottom of the boundary layer. The noisy values might be due to interpolation of the shape function profiles from the GOTM data. 
Based on hyperparameter sweep, we chose two hidden layers with 32 hidden nodes for $\mathcal{N}_1$, for which average correlation $\approx$ 0.9, and it is reasonably close to more expensive networks. For a deeper and wider network than 32 nodes, the average correlation score does not vary significantly, but the cost of using the network in an OGCM increases.

\begin{figure}[ht]
    \centering
    \includegraphics[width=\linewidth]{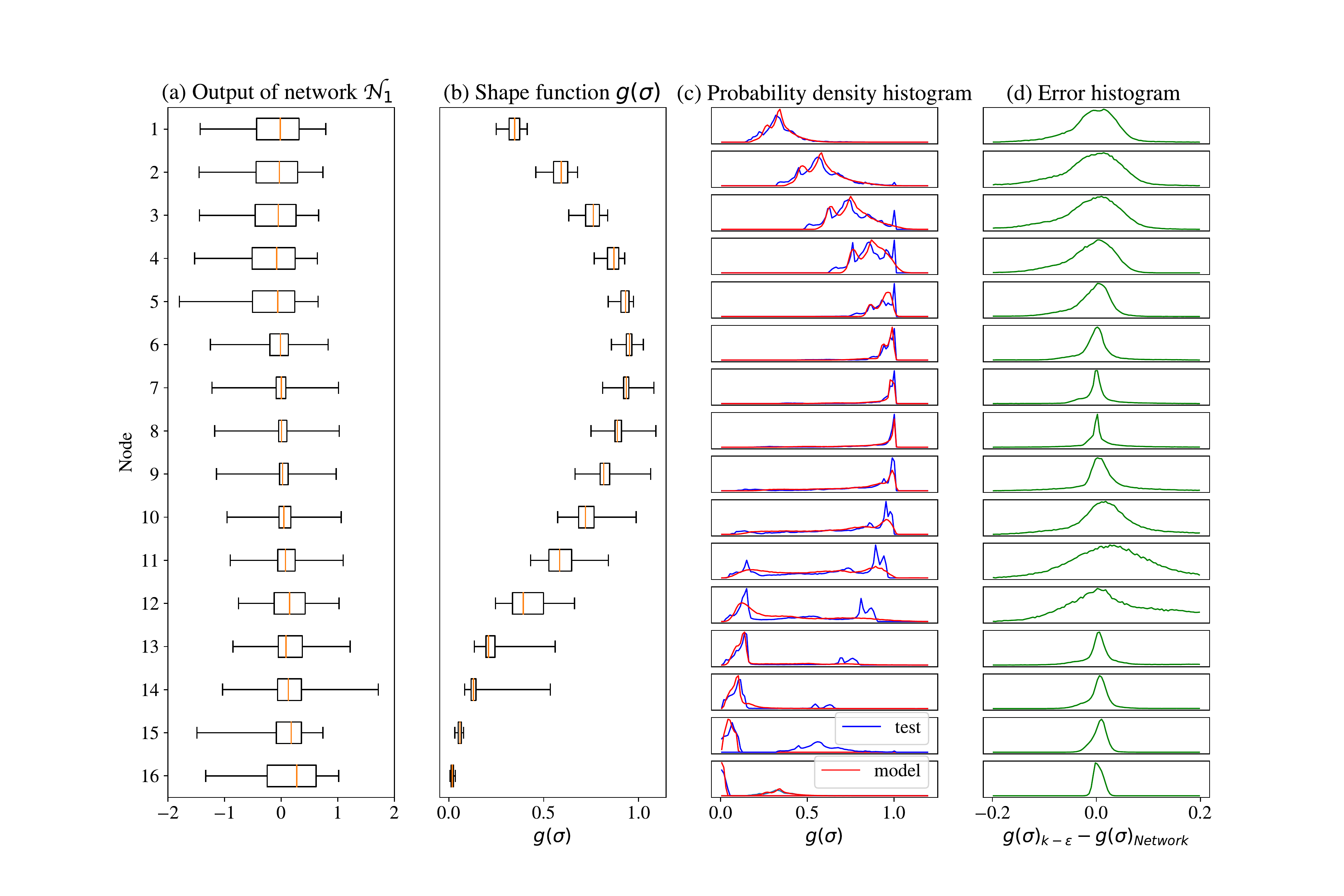}
    \caption{Performance of $\mathcal{N}_1$ for all the 16 sigma points. (a) Difference between network prediction and data. (b) Difference between network prediction and data in the physical space. Percentile ranges have been superimposed over the mean shape function from the training dataset.(c) Probability density histogram between network prediction and data. (d) Histogram of error defined by differences between the network prediction and data. For nodes 10, 11, and 12 networks  exhibits poor performance. This could be due to the strong multi-modal nature of data at those locations.}
    \label{fig:n1stats}
\end{figure}

Figure \ref{fig:r2scores} (b) shows the loss curves for training the network. Training loss (magenta) and validation loss (green) decrease with the training epoch. The validation loss is higher than the training loss, but both eventually plateau. The difference between validation and training loss, shown in blue, remains constant in later epochs, signifying when training should be stopped. The validation loss does not increase, ensuring that the network is not overfitting the training data. 
The performance of the network is further tested by comparing it with the validation data.  Strong agreement with validation data can be seen through the average correlation scores in Figure \ref{fig:r2scores}(a) and the error statistics in column (d) of Figure \ref{fig:n1stats}. 

Figure \ref{fig:n1stats} displays the performance of $\mathcal{N}_1$. The first column (a) shows the error statistics between the validation data and the network's prediction in the normalized space for each output node. The boxes show the interquantile range, while the whiskers show the 5$^{th}$ and 95$^{th}$ percentiles of the error. The second column (b) shows the same percentile range as in column (a) but in the physical space of $g(\sigma)$. The medians are superposed over the mean $g(\sigma)$ profile of the entire dataset. This helps to visualize the skill of $\mathcal{N}_1$ with respect to each $\sigma$ value.
Nodes 11 and 12 have a high error variance compared to other nodes. The error variances in column (a) are different from those in column (b) because the data have different variances along the nodes. The last node 16 has a high variance in (a) but because the $g(\sigma)$ values at that node are small, poor performance at that node does not penalize the overall performance of $\mathcal{N}_1$. Node 16 is in the transition layer, which may have a large gradient of the tracer that might amplify the error in diffusivity at node 16. However, implementing this version of the network in ePBL yields an acceptable improvement in overall performance, suggesting that the error in node 16 is acceptable. Sensitivity in the transition layer will be investigated in more detail in future work.
Column (c) has histogram plots of the validation data and its prediction. The network performs reasonably well and only shows inaccurate behavior when the data is multimodal. Column (d) shows the error histogram. The error has the highest variance at node 12, and is approximately Gaussian everywhere implying randomness.  

For the neural network $\mathcal{N}_1$, Figure \ref{fig:n1stats} shows the ability to predict the shape function offline. In general, the network shows high skill, as seen by the scores in Figure \ref{fig:r2scores}. The network $\mathcal{N}_1$ shows some inaccuracies in predicting multimodal distributions for output nodes 10-15. A single network predicts the value of the shape function at all the nodes, and it could compensate for the accuracy at one node over the other. Increasing the size of the network (i.e. number of layers and nodes in them) slightly reduces this error, but the cost of computation increases significantly with size rendering them unusable for longer time-scale simulations. %

$\mathcal{N}_1$ is trained in \textit{all} of the forcing regimes: surface heating, neutral, and convection. Perhaps, this adds a limitation to the network, which falls short of having very high skill for all the regimes. In our training experiments (not described in this article), training and predicting separately on the stable and unstable regimes gave higher skill than training on all regimes at once. Having two networks to predict $g(\sigma)$ alone could lead to higher skill without increasing the number of hyperparameters. This might be a cost-effective way to increase the overall accuracy of ePBL\_NN without expanding the size of the network. Increasing the number of hidden nodes in the hidden layers increases the cost of forward computation, while switching between the networks based on the forcing regime has a similar cost to using a single network. For simplicity, in this work we prefer to train all data using a single neural network and have not pursued this any further. %

We used the L1 loss function (mean absolute error) for training, as it gave better training performance than the L2  (root mean square error). We also increased the convergence of the network parameters (weight and biases) by tweaking the loss function. The loss values at nodes 8 to 13 were amplified by a factor of 100. This made the loss gradients steeper at the nodes that show the highest variance (seen in columns 1 and 2). This forces the network to put more weight on reducing errors on the nodes that are otherwise difficult to learn. The ADAM optimization algorithm \citep{adam} has been used to train the weights and biases of the network with a learning rate of $10^{-3}$.

\subsection{Training $\mathcal{N}_2$}\label{sec:N2_training}

The second neural network $\mathcal{N}_2$ as shown in Figure \ref{fig:neuralnetworks} predicts the characteristic velocity, $v_0$. Velocity is diagnosed from the training data using the following \emph{jugaad}: %
\begin{linenomath*}
\begin{equation}
    v_0 = \overline{\left( \frac{\text{max} (\kappa_{k\epsilon}(\sigma))}{h} \right)}
    \label{eq:v}
\end{equation} 
\end{linenomath*}
where the overbar denotes the average of all the values of $v_0$ for a set $(f, B_0, u_*)$. The spread of $\text{max} (\kappa_{k\epsilon}(\sigma))/h$ for a constant $(f, B_0, u_*)$ is small, and averaging assists the neural network in training to predict the mean value (see \ref{sec:appendixv0}).

\begin{figure}[ht]
    \centering
    \includegraphics[width=\columnwidth]{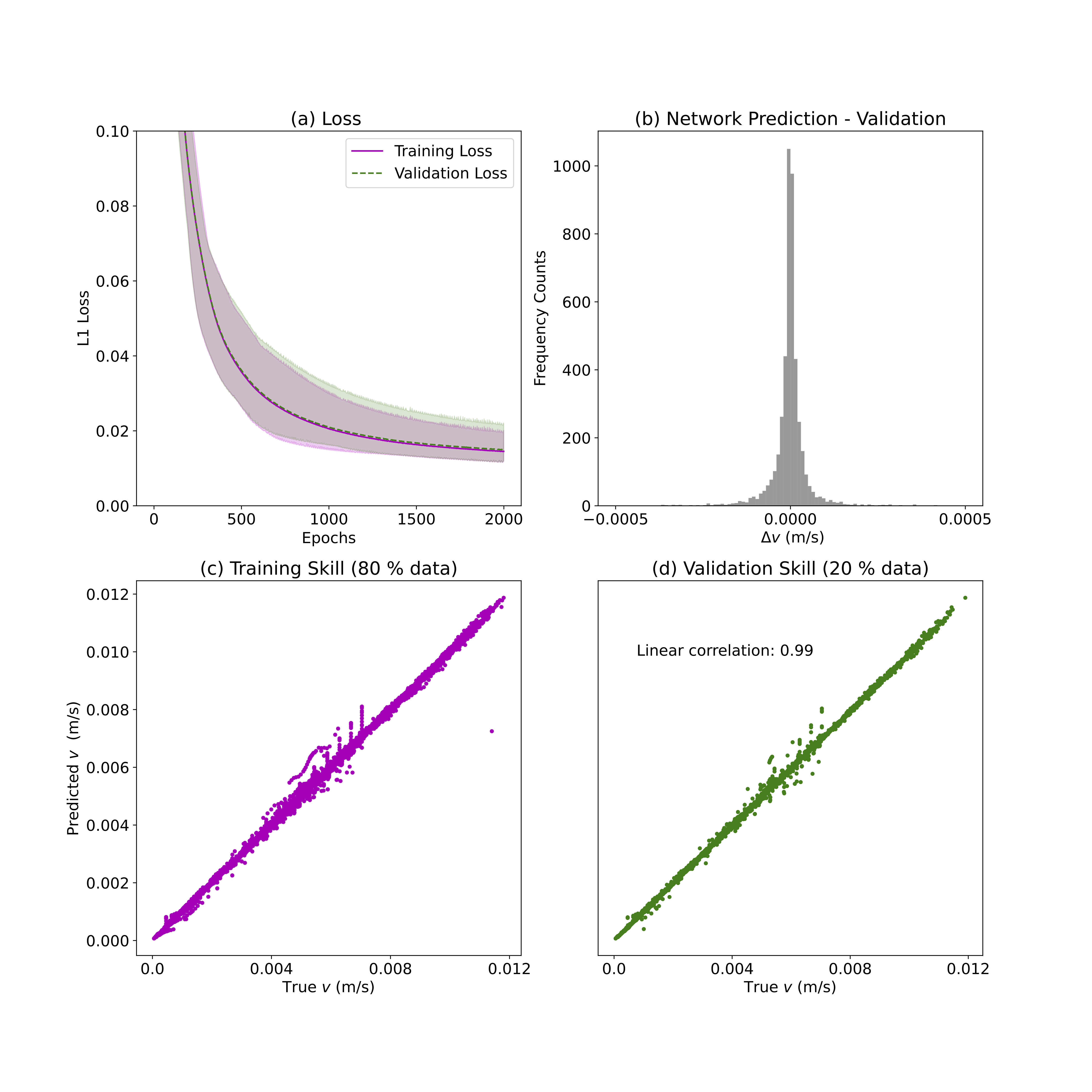}
    \caption{Performance of $\mathcal{N}_2$. (a) Loss curves. (b) Histogram of difference between network's prediction and data. (c) Predicted vs. true values for the training dataset. (d) Predicted vs. true values for the validation dataset. }
    \label{fig:n2train}
\end{figure}

Similarly to $\mathcal{N}_1$, network is trained on logarithm of $v_0$ and exponential function is used while inferring to ensure that the predicted $v_0$ is strictly positive. The data is divided into 80-20\% to train and test the performance of the network. As seen in Table \ref{table:table1}, the training data cover a wide range compared to that of $\mathcal{N}_1$, including extreme forcing conditions anticipated in a realistic OGCM. When the network sees conditions outside this range, the input is capped at the nearest extremum data point. This is to prevent the network from extrapolating, which is less skillful than interpolation. The trained network has high skill (linear correlation of 0.99) as seen in Figure \ref{fig:n2train}.

\section{Evaluating impacts in a prognostic OGCM}

Training, testing, and validation data provide one method for testing the network and its ability to reproduce training data. However, to fully test the network's potential for OGCM experiments the neural networks must also be implemented in free-running, prognostic models. %
Our implementation does not cause simulation to fail due to any spurious effects or instabilities which is a known problem with implementing neural networks in a GCM (see \cite{Brenowitz2020} and references therein). Stability might result because we implement neural networks as a component within the existing ePBL framework. We demonstrate the success of our implementation using both free running column model experiments and forced ice-ocean global OGCM climate model experiments.

\subsection{Implementation of neural networks in MOM6}
\label{sec:implement}

We now describe the implementation of our networks in the MOM6 ocean model. 
The weights and biases of the network are generated offline and stored in NetCDF files. Feedforward (inference) calculation of the network involves matrix multiplications and activation functions. These have been coded as subroutines in MOM6's vertical mixing module (ePBL). A flag activates the neural networks to predict $g(\sigma)$ and $v_0$. All inputs to the network are readily available within the ePBL module. The neural networks require the depth of the boundary layer $h$, which is provided by the ePBL scheme as described in \cite{Reichl2018}. The neural networks function alongside the algorithm by which ePBL derives $h$ and therefore they do not interfere with any energy constraints set by the original scheme. Additionally, in MOM6, the diffusivity derived from ePBL and the neural network subroutines is passed to a main diabatic mixing module which combines diffusivities from various mixing parameterizations (such as \cite{JHL}) within MOM6. More details can be found in \cite{Reichl2018}.  
$g(\sigma)$ is obtained at 16 points between the surface and $h$. $g(\sigma=0)=0$ to satisfy zero diffusivity at the surface. At $h$, $g(\sigma=1)$ is set as a small number by assuming $g(\sigma=1) = c \cdot g(\sigma=16/17) $, where $c$ is a small positive constant set as 0.1. GCM and single column runs were found to be insensitive to small and non-zero values of $c$.\\
Shape function on $\sigma$ is converted to the model's vertical grid by linear interpolation. The use of the sigma coordinate makes our scheme grid independent of the vertical coordinate. The shape function on the model grid is multiplied by $v_0 \cdot h$ according to Equation \ref{eq:knn} to recover the diffusivity profile of the $k-\epsilon$ model. The subroutines pass on the diffusivity profile to the ePBL module. In MOM6, there are other parameterizations active along with ePBL to incorporate strong shearing regions found at the equator and also that handle background diffusivity.
Both networks $\mathcal{N}_1$ and $\mathcal{N}_2$ are shallow, as they have two hidden layers with 32 nodes in each. OM4 model with ePBL\_NN requires $\approx$ 5-10\% more runtime than ePBL.
This cost may not warrant a need for GPUs to speed-up the inference in this version of the scheme, but this option will be explored in the future. 

The inputs to the neural network are also capped inside the subroutine to ensure the networks do not make predictions outside their training range. For $\mathcal{N}_1$, if any of the inputs $(f, B_0, u_*, h)$ are outside the known range, then the subroutine limits the inputs and changes them to the nearest point in the four-dimensional hypercube formed by the four inputs. Our training data covers a reasonable space of the forcing parameter regime as observed among realistic present conditions (as it will be applied in this study).  Data points outside the range are less probable, allowing the network to perform effectively for nearly all of the tested forcing conditions (see \ref{sec:appendixhx}). Capping the inputs prevents the network's output from being unphysical.  If the network is applied for simulations in substantially different climate regimes (e.g., paleoclimate or for other planetary bodies) the training data could be enhanced. 
If the network receives inputs outside the known range, the shape function can have spurious values with irregular vertical structure. Capping the inputs ensures that this spurious behavior is prevented. The training on logarithm and using exponential function while inferring described in the earlier sections prevents non-positive behavior for both $\mathcal{N}_1$ and $\mathcal{N}_2$.

\subsection{Single Column Model Results}
\label{sec:SCM}

We compare three schemes to examine the performance of the network in single column model: GOTM $k-\epsilon$, ePBL, and ePBL\_NN. MOM6 in single column configuration \citep[as in][]{Reichl2918} is used to run ePBL and ePBL\_NN, while GOTM is used for the $k-\epsilon$ experiments. The column models are forced at the surface grid interface with constant buoyancy forcing (surface heating of 50 W/m$^2$) and constant wind surface stress (0.2 N/m$^2$). Stratification is constant throughout the column in the initial conditions. To have the same entrainment in all the three cases, the m$^*$ value is diagnosed from the $k-\epsilon$ output and imposed in MOM6. The quantity m$^*$ is the non-dimensional integral of the entrainment flux and is given by $ \int_{-h}^{0} \min (0, \overline{w' b'}) dz = m^* u^{*3}$ for surface heating conditions. In \cite{Reichl2018}, m$^*$ has been parameterized using a function $G$ as in Eq. \ref{eq:w'b'}. Instead of using the parameterized m$^*$ from \cite{Reichl2018, Reichl_Li_2019}, we use a diagnosed and time varying m$^*$ from $k-\epsilon$ to perform a controlled comparison with identical forcing conditions. This prevents deficiencies in the parameterized m$^*$ from causing any disagreements between MOM6 and GOTM. By matching the surface forcing and integral of the entrainment flux, the differences between all the three cases can only be due to diffusivities. 

\begin{figure}[h!]
\centering
\includegraphics[width=1.0\textwidth]{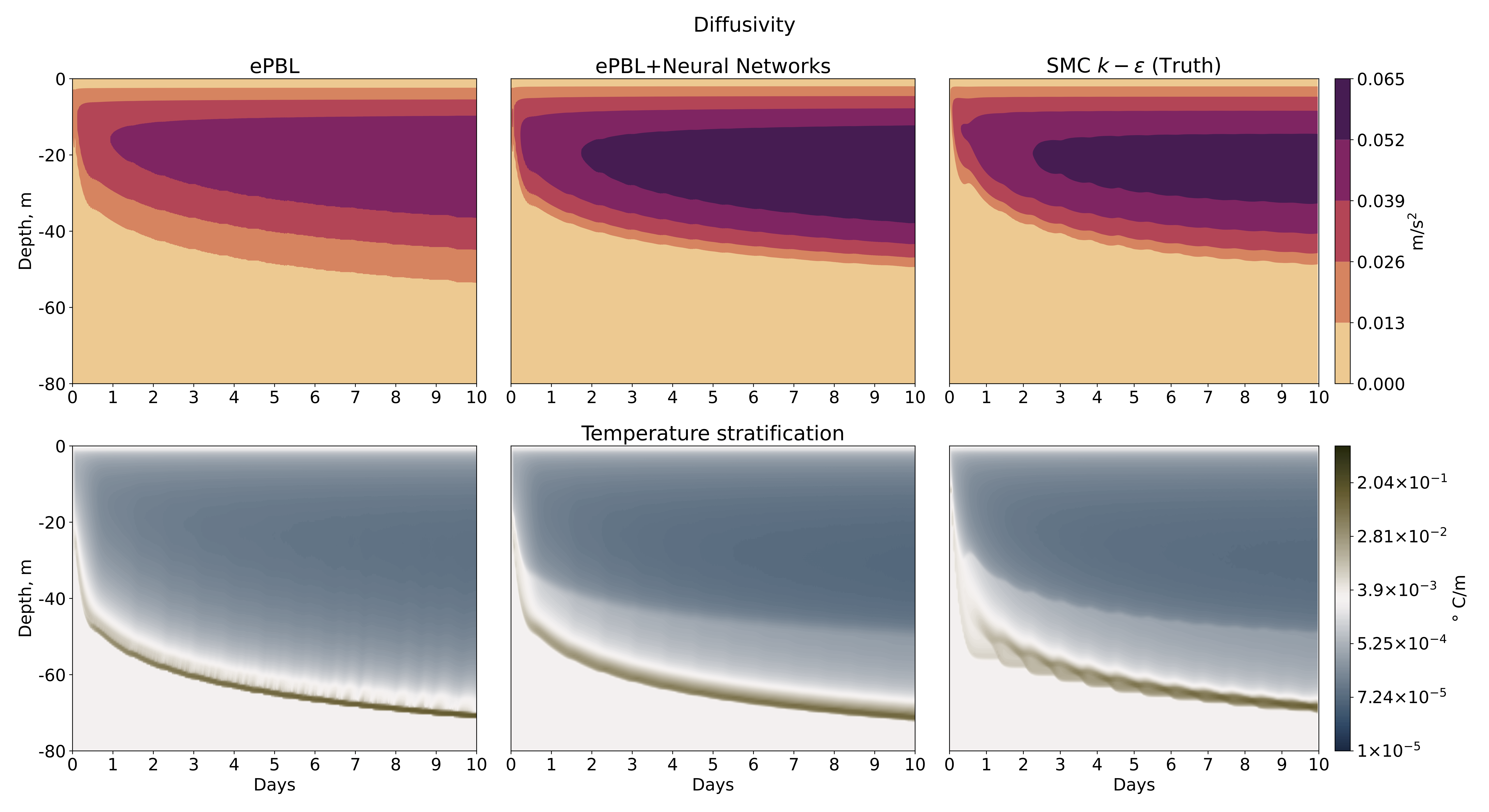}
\caption{Time series comparison for single column model configuration. Latitude set to 40 \textdegree, surface heat flux is 50 W/m$^2$, and wind stress is 0.2 N/m$^2$.  The upper row compares diffusivity and the bottom row compares stratification. In both the cases, ePBL\_NN is in better agreement to the second moment closure scheme $k-\epsilon$ than ePBL.}
\label{fig:surfheat_hov}
\end{figure}

\begin{figure}[h!]
\centering
\includegraphics[width=1.0\textwidth]{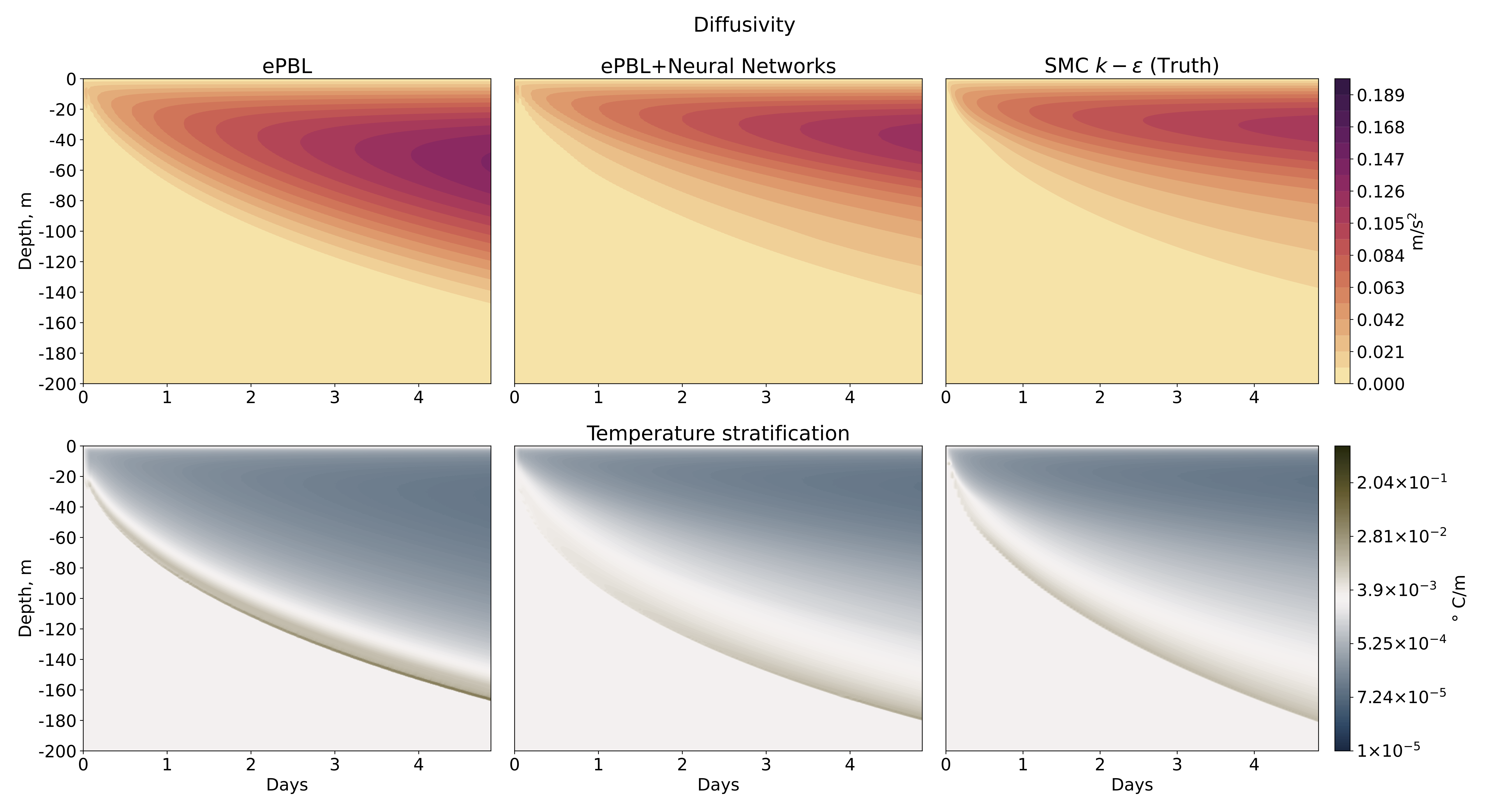}
\caption{Time series comparison for single column model configuration. Latitude set to 1\textdegree, surface heat flux is 50 W/m$^2$, and wind stress is 0.2 N/m$^2$.  The upper row compares diffusivity and the bottom row compares stratification. In both the cases, ePBL\_NN is in better agreement to the second moment closure scheme $k-\epsilon$ than ePBL.}
\label{fig:surfheat_hov_lat1}
\end{figure}

Two latitudes are compared: Latitude 40\textdegree  (Figure \ref{fig:surfheat_hov}) and 1\textdegree (Figure \ref{fig:surfheat_hov_lat1}). The figures show the time series of diffusivity and temperature stratification. For both latitudes, the diffusivity and stratification in ePBL\_NN are in closer agreement with the $k-\epsilon$ model than the original ePBL model, showing the ability of the neural networks to match $k-\epsilon$.
ePBL\_NN has a diffusivity profile closer to $k-\epsilon$ than ePBL throughout the OSBL. In $k-\epsilon$ (SG), the turbulent diffusivity is computed from the simulated TKE and turbulent length scale, using stability functions that relate the Prandtl number to the Richardson number \citep{schumann1995}. The neural networks have ``learned" those relationships (without direct knowledge of either parameter) that set the structure of diffusivity and hence show high skill in predicting the profile. 

The upper $\approx 20 \%$ of the diffusivity profile is able to learn traditional constraints, such as the law of the wall scaling, since they are features of the training data. The bottom $\approx 40 \%$ of the OSBL shows more variability and is an important region for the entrainment process. In deepening of the boundary layer, the entrainment process mixes the higher density water masses (usually cold) from below the mixed layer with the lower density mixed layer above it (usually warmer). Outside of the polar regions, this process cools the mixed layer along with the sea surface temperature (and warms the interior) and has implications for ocean-atmosphere energy exchange and feedbacks. 

\subsection{Ice-Ocean JRA Forced Model Results}
\label{sec:JRA-forced}

We next tested the ePBL\_NN scheme using the GFDL's OM4.0 ocean/sea ice model. The model has a nominal 1/4 degree resolution and is forced using the JRA forcing product \citep{jra55}. JRA forced simulations constrain the atmospheric fields that force the ocean model with the observed/reanalysis atmospheric data. This is different from the atmosphere-ocean coupled model as there is no feedback from the ocean response to the atmosphere. However, this approach is beneficial for testing parameterizations since two experiments can be more carefully compared without considering the complications of those feedbacks.  Future work will examine the performance of these schemes in fully coupled climate models. 

Two sets of OGCM experiments have been performed: one using the ePBL scheme as a control run (e.g., as described by \cite{Adcroft2019}) and the second with the neural networks active to replace the shape function and velocity scale in ePBL. The simulations were performed for a period of 1958 to 2017.

In this study, we compare the two runs with observations to analyze the impact on: (1) Ocean heat uptake, (2) Sea surface temperature, (3) Mixed layer depth, and (4) Upper ocean temperature stratification in the Tropical Pacific.
For sea surface temperature, data from the World Ocean Atlas (WOA) \citep{woa_data,woa_data_temp} has been used to compare the two schemes. For the subsurface comparison: mixed layer depth and stratification, ARGO float measurements have been utilized \citep{ARGO}. 

\subsubsection{Ocean Heat uptake and Sea Surface Temperature Comparison}

\begin{figure}[h]
    \centering
    \includegraphics[width=\textwidth]{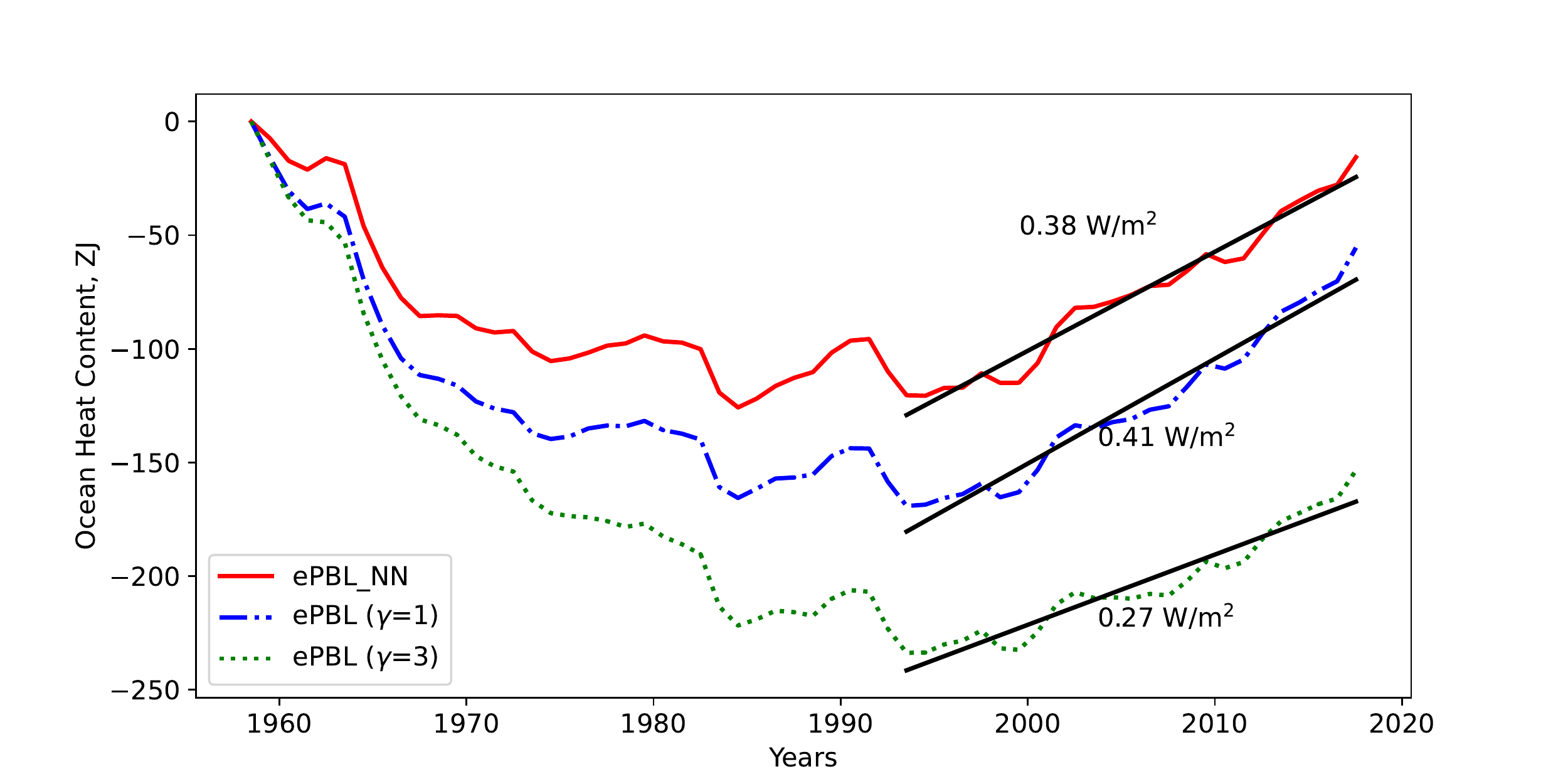}
    \caption{Total ocean heat content compared between ePBL and ePBL\_NN for the duration of 1958-2017. For ePBL, $\gamma$ from Eq. \ref{eq:lzgamma} has been set to 1 and 3. For ePBL\_NN, the tunable parameter $\gamma$ does not exist. We can observe that the ocean's total heat content is sensitive to the vertical diffusivity set by the OSBL mixing scheme. ePBL\_NN replaces ad-hoc diffusivity of ePBL with a physics informed data-driven neural network.}
    \label{fig:thetaoga}
\end{figure}

Figure \ref{fig:thetaoga} shows the global ocean heat content for the three runs: one with ePBL\_NN shown in red-solid line, and the other two with ePBL by setting $\gamma$ from Equation \ref{eq:lzgamma} as 1 and 3 shown as blue-dashed line and green-dotted line respectively. ePBL\_NN shows more heat uptake than the original scheme, and rate of warming is between ePBL runs with $\gamma$ set as 1 or 3. This highlights the sensitivity of the total ocean heat content to the shape function and to boundary layer mixing schemes.

\begin{figure}[h]
\centering
\includegraphics[width=\textwidth]{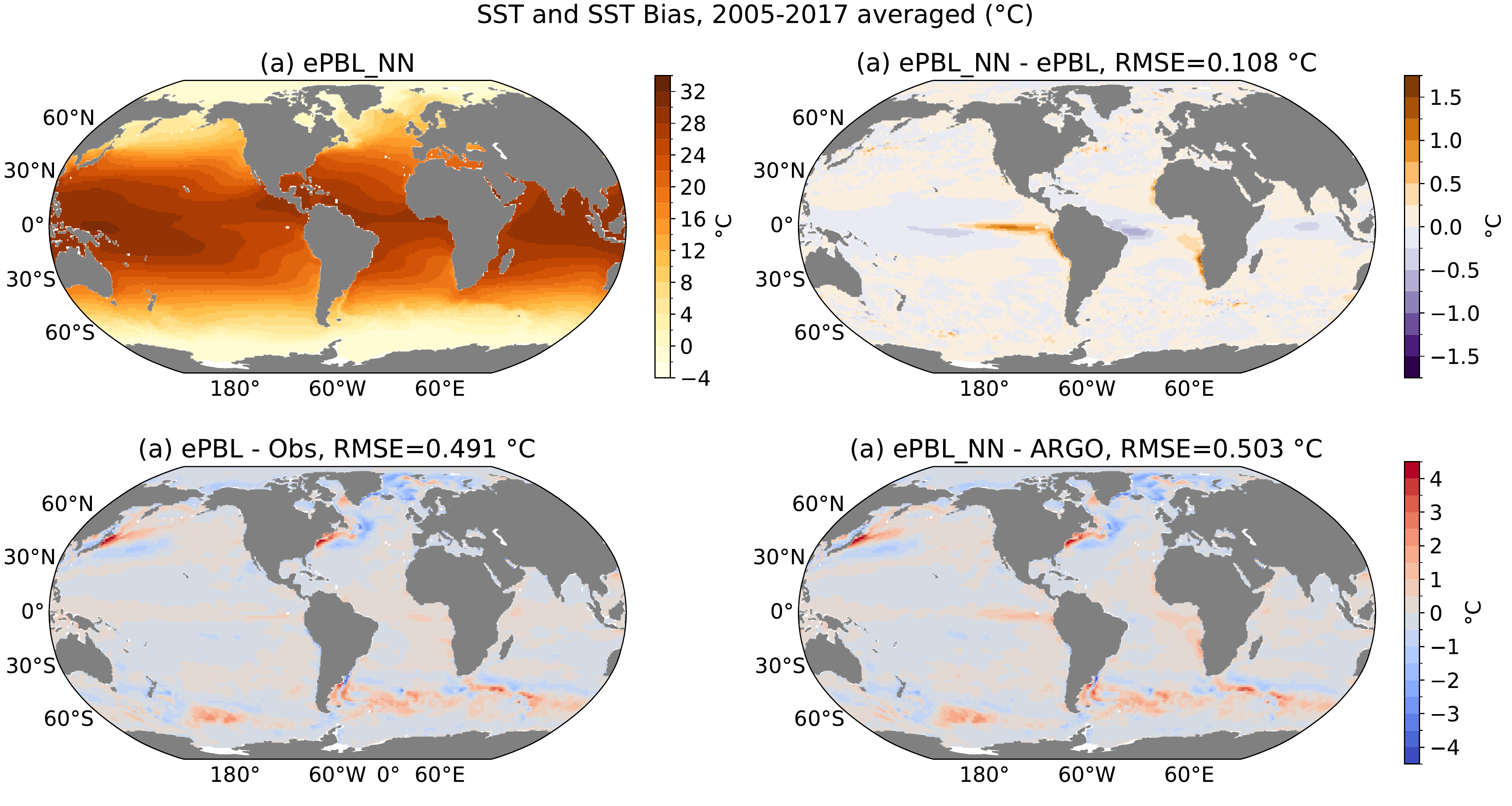}
\caption{Sea surface temperature and biases. (a) SST from model runs using ePBL\_NN. (b) SST difference between ePBL and ePBL\_NN. (c) SST bias between ePBL and observations (Obs). (d) SST bias between ePBL\_NN and observations. Bias plots also show mean and standard deviation. Observations are from World Ocean Atlas dataset \citep{woa_data_temp}. SST biases are similar for ePBL and ePBL\_NN. At the equatorial region, ePBL\_NN shows slightly colder bias than ePBL.}
\label{fig:sst}
\end{figure}

Figure \ref{fig:sst} shows the SST bias averaged over the years 2003-2017 for each 1\textdegree grid point.  SST biases are similar in the two runs with minimal differences, which is expected since the atmospheric fields are prescribed and not coupled. SST around the eastern Pacific and Atlantic equatorial regions shows a slightly warmer bias for the ePBL\_NN run than for the ePBL. In the Indian ocean, the bias is slightly colder. The SST bias in the Gulf Stream and Kuroshio current is slightly warmer in ePBL\_NN by about 0.5 \textdegree C. The response of the SST to ePBL\_NN in the boundary current regions indicates that changes in the vertical viscosity or diffusivity also impacts the circulation in certain regions.

Changes in the patterns of SST can be due to changes in the mixed layer depth and the surface heat fluxes. The heat fluxes are computed as a function of SST, surface ocean velocity and ice cover as stated in \cite{Adcroft2019, Griffies2016omip}.

\subsubsection{Mixed Layer Depth Comparison}

\begin{figure}[h]  %
\centering
\includegraphics[width=\textwidth]{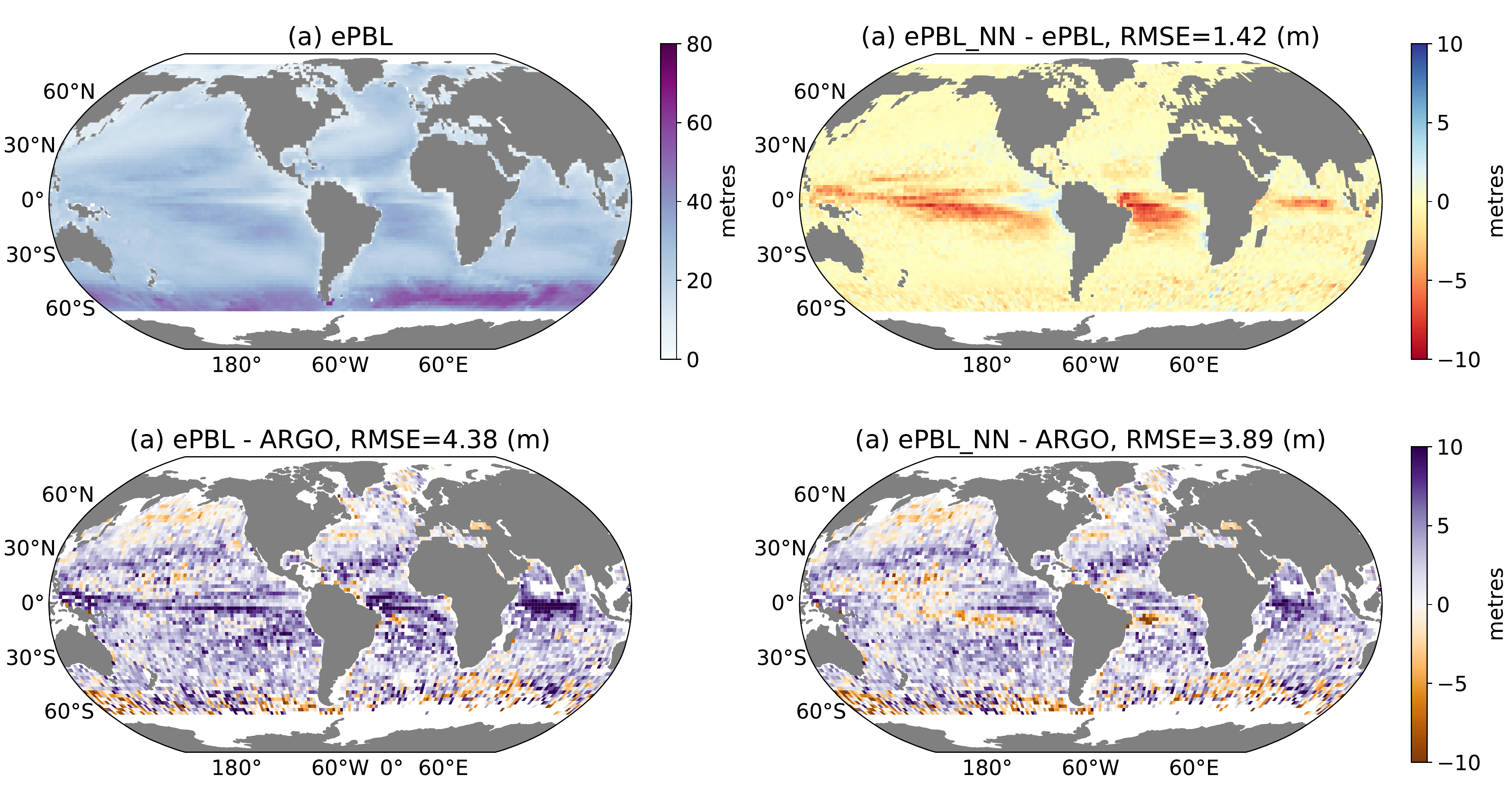}
\caption{Summer time (shallow) mixed layer depth biases using the Potential anomaly criterion of \cite{ReichlMLD}. (a) MLD from ePBL, (b) Difference of MLD between ePBL and ePBL\_NN. (c) Bias of ePBL with respect to ARGO data, (d) Bias of ePBL\_NN with respect to ARGO data. We can notice the bias reduction from (c) to (d).}
\label{fig:summer_mld_Pe}
\end{figure}

\begin{figure}[h]  %
\centering
\includegraphics[width=\textwidth]{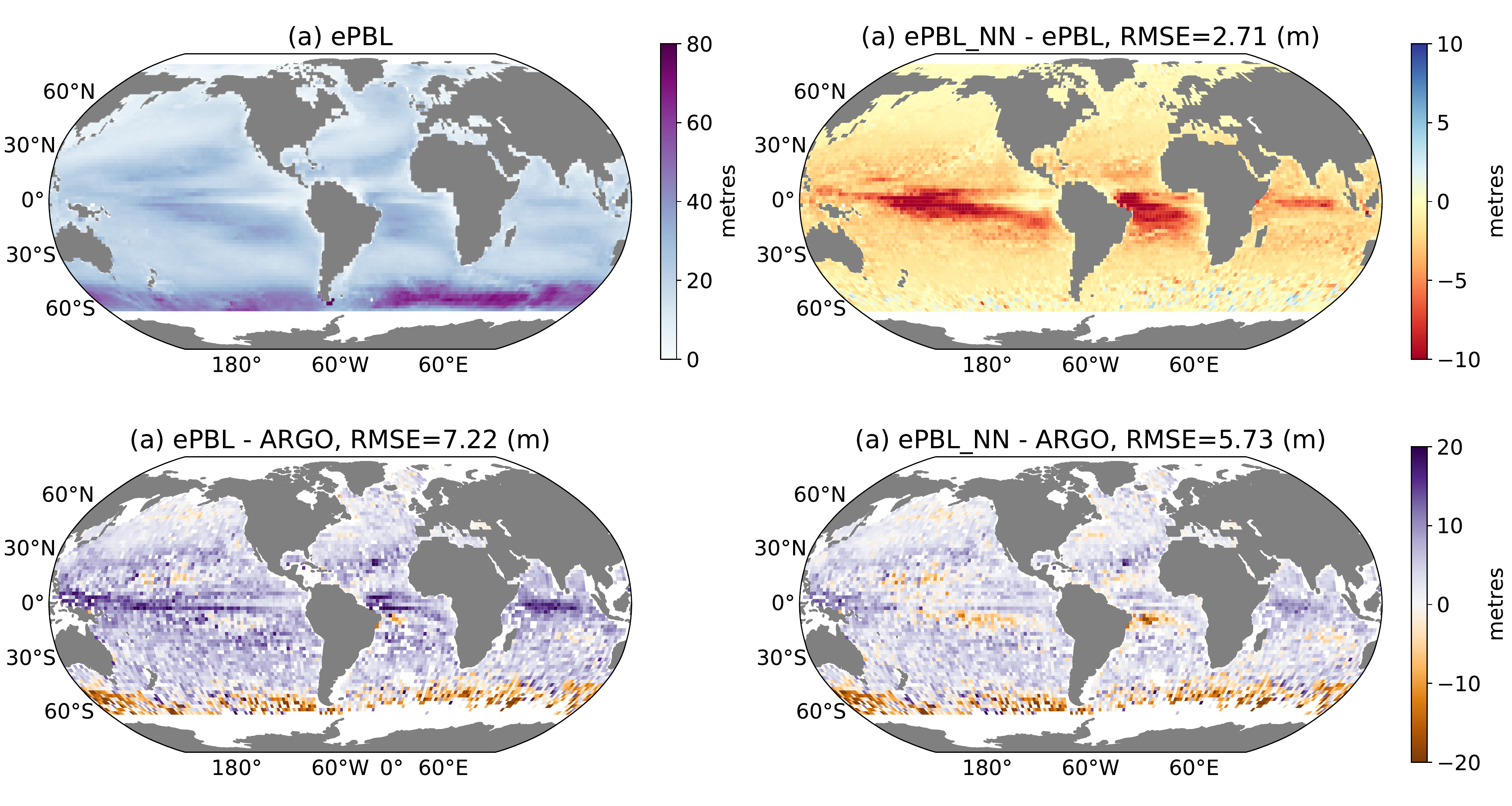}
\caption{Summer time (shallow) mixed layer depth biases using the density criterion of \cite{MLD03}. (a) MLD from ePBL, (b) Difference of MLD between ePBL and ePBL\_NN. (c) Bias of ePBL with respect to ARGO data, (d) Bias of ePBL\_NN with respect to ARGO data. We can notice the bias reduction from (c) to (d) and is consistent with that observed in Figure \ref{fig:summer_mld_Pe}}
\label{fig:summer_mld_De}
\end{figure}

Summer and winter mixed layer depths (MLD) are compared, a metric usually used to indicate the depth at which atmospheric influences are directly felt in the ocean. Here, winter (summer) mixed layer depth is the maximum (minimum) of the monthly averaged MLDs for each grid point over the period 2003-2017.  The MLD depends on the definition, and we evaluate it using two criteria: \cite{ReichlMLD} and \cite{MLD03}. The criterion from \cite{MLD03} uses a threshhold potential density of 0.03 kg/m$^3$ whereas \cite{ReichlMLD} uses a threshold potential energy anamoly of 25 J/m$^2$ to define the MLD. Figures \ref{fig:summer_mld_Pe} and \ref{fig:summer_mld_De} show the MLD using the potential energy anomaly criterion and the potential density respectively. 

\begin{figure}[h]  %
\centering
\includegraphics[width=\textwidth]{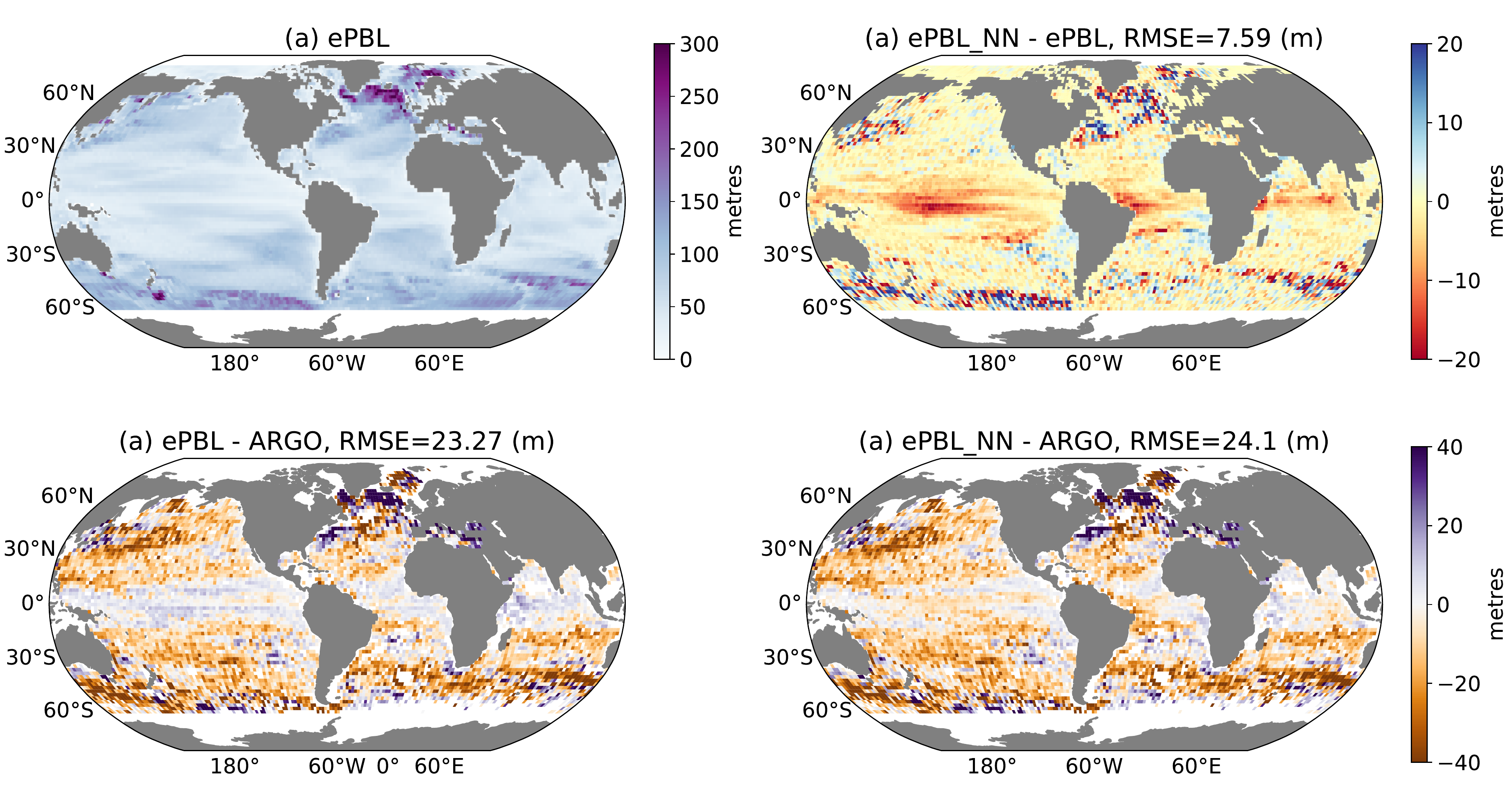}
\caption{Winter time (deep) mixed layer depth biases using the Potential anomaly criterion of \cite{ReichlMLD}. (a) MLD from ePBL, (b) Difference of MLD between ePBL and ePBL\_NN. (c) Bias of ePBL with respect to ARGO data, (d) Bias of ePBL\_NN with respect to ARGO data. Biases are similar in (c) and (d) and ePBL\_NN does not significantly worsen any biases. }
\label{fig:winter_mld_Pe}
\end{figure}

\begin{figure}[h]  %
\centering
\includegraphics[width=\textwidth]{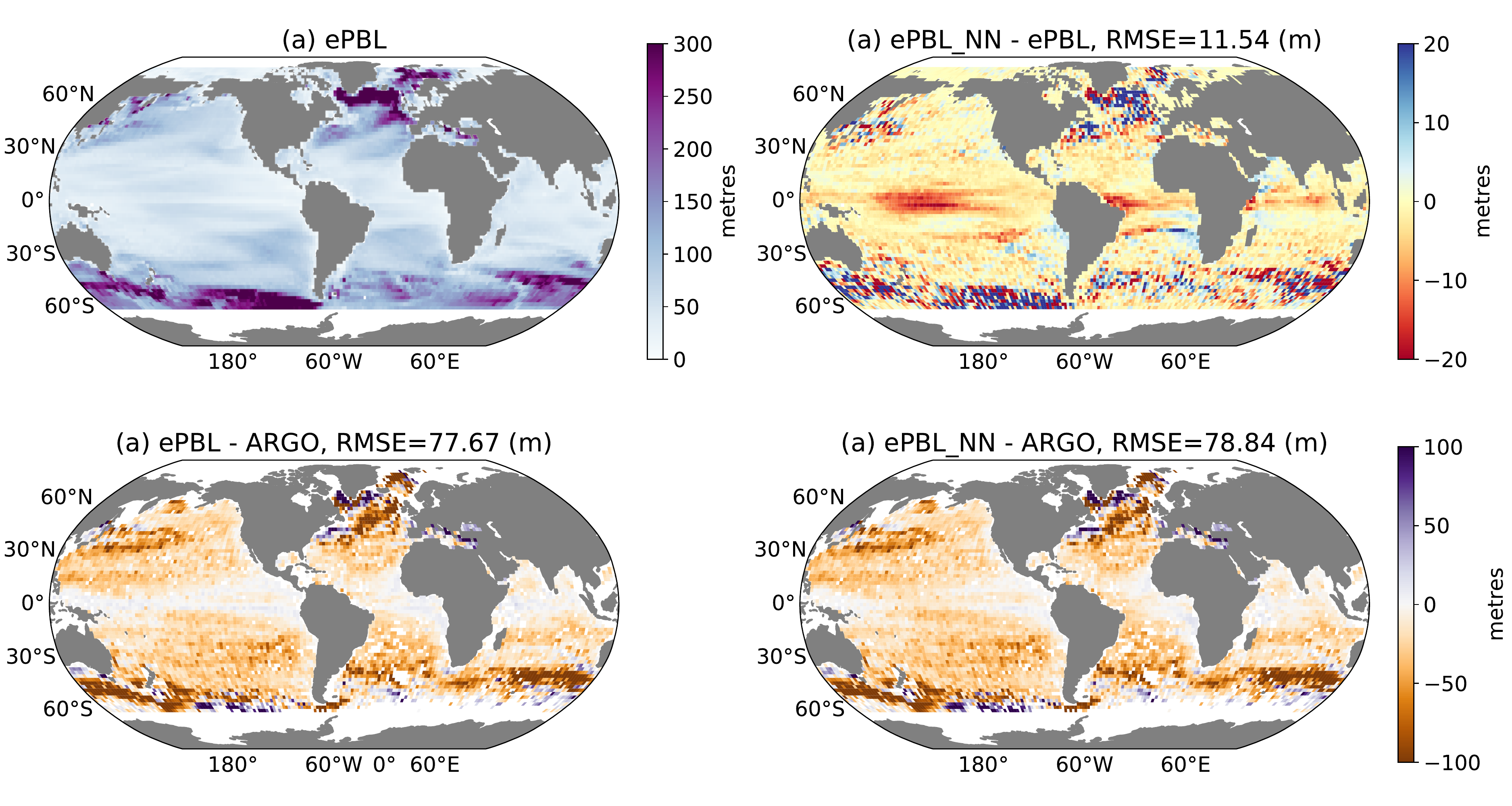}
\caption{Winter time (deep) mixed layer depth biases using the density criterion of \cite{MLD03}.  (a) MLD from ePBL, (b) Difference of MLD between ePBL and ePBL\_NN. (c) Bias of ePBL with respect to ARGO data, (d) Bias of ePBL\_NN with respect to ARGO data. Biases are similar in (c) and (d) and ePBL\_NN does not significantly worsen any biases. This observation is consistent with that observed in Figure \ref{fig:winter_mld_Pe}}
\label{fig:winter_mld_De}
\end{figure}

Figures \ref{fig:summer_mld_Pe}, \ref{fig:summer_mld_De} show summer time MLD. The summer time MLD bias has reduced significantly in ePBL\_NN as compared to ePBL. The average bias reduced from 7.22 m to 5.73 m as seen in Figure \ref{fig:summer_mld_De}. Between -20\textdegree $ $ to 20\textdegree $ $  latitude, the average root mean square error (RMSE) for MLD bias in ePBL was about 7.94 m. In ePBL\_NN, the bias was reduced to 5.18 m. We have shown the latitude dependency of RMSE between model and observations in the supplementary section (see Figure S1).
The ePBL\_NN scheme performs better under stable surface heating conditions than the ePBL scheme. The shallow MLD bias reduction has implications for equatorial oceanic regions and its effect on large-scale ocean-atmosphere feedbacks \citep{Adcroft2019}. %
Winter MLD biases (Figures \ref{fig:winter_mld_Pe} and \ref{fig:winter_mld_De}) are very similar for both runs. %
The ePBL\_NN predicts diffusivity close to a second moment scheme but does not significantly impact the winter time bias simulated by the model with the original ePBL scheme. This is likely because other model physics and factors can  dominate in setting the deep convective mixed layers and water properties. %

Although ePBL\_NN has been trained on all the regimes including surface cooling, a different scheme or process might be compensating  the effects of improved diffusivity. This could also be due to higher sensitivity of shallow mixed layers to changes in surface forcing than deep mixed layers. For shallow mixed layer depth, any perturbations in the atmospheric forcing will reach the base of the boundary layer quicker than it would reach in deeper layers. In \cite{Reichl2018}, the rate of conversion of turbulent kinetic energy to potential energy within the boundary layer (left hand side in Equation \ref{eq:ePBL_k}) uses a parameterization that depends on $h$. Changing the diffusivity can alter $h$ which in turn modulates the rate of energy conversion. This can lead to changes in the MLD.

The MLD evaluated using the criterion of \citep{MLD03} shows the same qualitative results as described above. The winter time MLD biases are similar for both runs. The summer time MLD bias shows a further reduction when evaluated using \citep{MLD03} than with \citep{ReichlMLD}. It is not unusual to get different values of mixed layer depths using different definitions. For both definitions, winter biases in ePBL\_NN are not worsened. Qualitative agreement of the reduction in summer bias in ePBL\_NN using two different criteria provides strong evidence of ePBL\_NN performing better than ePBL in terms of MLD bias reduction under fixed atmospheric forcing conditions.

\clearpage

\subsubsection{Comparison of Upper Ocean Stratification in the Tropical Pacific}

\begin{figure}[h!]
\centering
\includegraphics[width=1.0\textwidth]{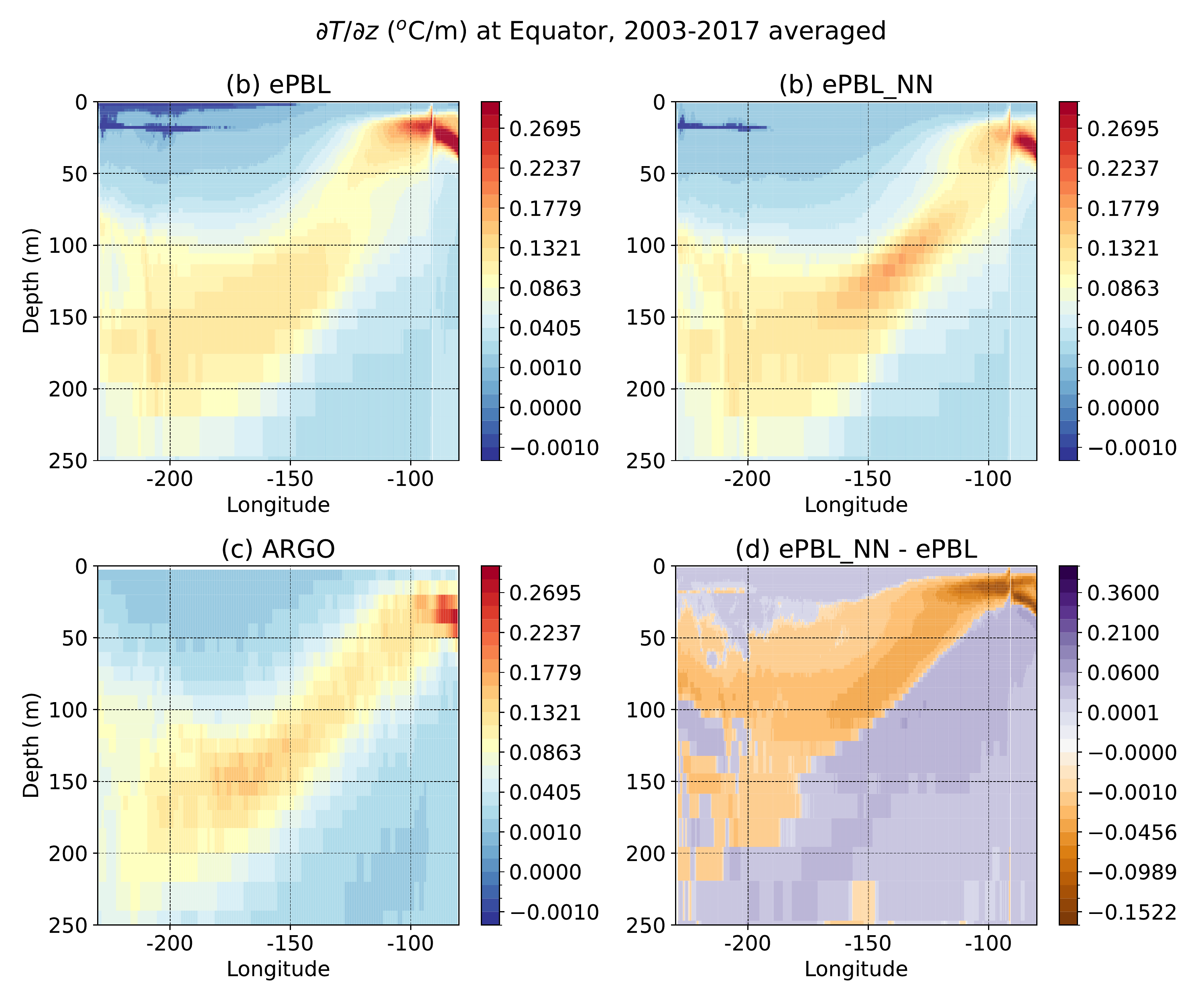}
\caption{Temperature stratification at a transect along the equator in the Pacific ocean. (a) ePBL output, control run. (b) ePBL\_NN output. (c) ARGO data (d) ePBL - ePBL\_NN. Note that (b) is closer to (c) than (a). ePBL\_NN has been instrumental in enhancing the stratification of the upper ocean.}
\label{fig:dtdz}
\end{figure}

The final comparison we use to assess the impact of the neural network diffusivities on the model result is the upper ocean temperature stratification ($\partial \Theta/\partial z$, where $\Theta$ is the potential temperature) in the Equatorial Pacific region. The thermocline in the Equatorial Pacific region plays an important role in ENSO dynamics with implications for the Earth's climate system [for e.g.] \citep{FeiFei}.  The temperature stratification is shown for a vertical cross section along the equator spanning -220\textdegree $ $ to -80\textdegree $ $ E. 
Figure \ref{fig:dtdz} (c) shows the observational data from ARGO floats \citep{Roemmich:2009}. Figure \ref{fig:dtdz} (b) is the $\partial \Theta /\partial z$ from the original ePBL and shows lower stratification as compared to ARGO observations. The $\partial \Theta/\partial z$ from ePBL\_NN, as seen in Figure \ref{fig:dtdz}(a), shows significant improvements in the stratification of the upper $\approx$ 50 m of the ocean. Stratification in ePBL\_NN is closer to ARGO data in the equatorial region of the Pacific ocean. The neural network predicted diffusivities help to increase the stratitication and make it closer to observations than the simulation with the original ePBL with the ad hoc shape function for diffusivity.

Stratification acts as a barrier to mixing, this warrants further investigation into how ePBL\_NN changes transport pathways of heat through the OSBL in different regions of the world's oceans. Overall, the MLD bias is reduced, and stratification has improved for the upper $\approx$ 50 m. suggesting that ePBL\_NN works to fix these two biases in conjunction.

\section{Concluding Remarks}

\subsection{Summary}

In this study, we apply neural networks to improve the parameterization of the vertical diffusivity in the ocean surface boundary layer. The data used to train the neural networks is obtained using second-moment closure simulations by running single-column model under various forcing scenarios, spanning the possible range of present-day and future conditions. 
The neural networks are implemented within the existing physics-based parameterization from the Energetic Planetary Boundary Layer (ePBL) framework of \cite{Reichl2018}.
The neural networks augment the method to determine the vertical diffusivity in the ePBL scheme with data-driven relations but maintain the physically motivated energetic constraints on mixing from the original scheme.
A benefit of our approach is that it yields a stable implementation in the OGCM (MOM6). This enables us to perform decade-scale simulations spanning 1958 to 2017.%

Atmospherically forced Ice-Ocean experiments using the GFDL's MOM6 1/4\textdegree model suggest an overall improved performance due to the enhancements in ePBL\_NN relative to the original scheme. There is a reduction in biases of summer-time mixed layer depths and no exacerbation of the winter-time biases compared to ePBL. The stratification of the upper ocean in the tropical pacific shows improvements in the thermocline compared to the ARGO float observations. This analysis indicates that the resulting scheme is suitable for implementation in future OGCM configurations and experiments and is expected to reduce biases in climate simulations.  Further analysis using a wider range of diagnostics in additional model configurations will be particularly beneficial.

The ePBL framework is already optimized for GCMs, providing larger time stepping capabilities ( $\approx$ $O(1)$ hr) and ePBL\_NN leverages these advances with improved diffusivity profiles. It is computationally expensive to run a second-moment scheme in a GCM due to time-stepping restrictions ( $\approx$ $O(10-100)$ s), but ePBL\_NN can yield eddy diffusivity profiles more consistent with a SMC within the original ePBL framework. This is significant for GCMs as we are achieving closer results to a model having a second-order turbulence closure scheme, but able to maintain coarse resolutions and long time steps needed for climate scale simulations. We note that the longer implicit time step used in the numerics of ePBL \cite[see][]{Reichl2018} can lead to a smoothing effect which can complicate resolving small-scale structure, but we observe that the large-scale evolution is tracked accurately.

While the results of this work are promising, numerous aspects remain important for future work.  
For example:
\begin{enumerate}
\item ePBL, ePBL\_NN, and the SMC considered here assume downgradient diffusion and hence have no nonlocal flux terms. The representation of nonlocal fluxes could improve the scheme further and potentially affect convective regions and Langmuir turbulence \cite[e.g.][]{Chor2021}.  In this application we do not explicitly consider the impact of Langmuir turbulence within ePBL (though it is part of setting the energy available for entrainment, see \cite{Reichl_Li_2019}). 
\item The neural networks can be made larger to capture more complex relationships in the data by increasing the number of hyper-parameters (hidden nodes).  In this work we chose small networks for initial investigation. The successful use of small neural networks as efficient surrogate models of SMCs proves that we can replicate the behavior of complex models with high fidelity. Increasing the network size will be explored in the future and will likely require using graphical processing units (GPUs) for implementation in OGCM \citep{Zhang2023}. %

\item The performance of the modified vertical mixing scheme in a coupled model (atmosphere-ocean-ice) may not show the same impact on model bias as observed in this forced ocean-ice model.  The atmosphere-ocean feedbacks will require exploration in future work. 
\item Improving the representation of the diffusivity profile has implications for many quantities that have gradients within the boundary layer.  For example, changing the diffusivity of nutrients within the euphotic layer has implications for biogeochemical processes such as primary production.  The implications of improved diffusivity for ecological modeling will be explored in future work.
\item Finally, we have trained on one SMC, the $k-\epsilon$ model with stability functions following \cite{schumann1995}. This parameterization was chosen for consistency with \cite{Reichl2018}, but alternative SMC models may yield different results. In future work this process will be repeated with different SMC schemes to understand the influence of SMC diffusivities on the performance of GCMs. One disadvantage is that SMCs have been assumed to be the ``truth" but it might lack realism and hence future work will focus on including data from LES studies and observations.  
\end{enumerate}

\subsection{Applications for first order ocean surface boundary layer parameterizations}

One key achievement of this work is that it establishes a relationship between the shape function of upper ocean vertical mixing and the forcing parameters. Previous work in similar first-order upper ocean mixing parameterizations assumes that the shape function is fixed, or was set by ad-hoc approximations. This work further suggests that models that consider this variation in the shape function are more skillful at simulating upper ocean stratification and ocean mixed layers.  The physics-informed functions (networks) developed in this work for determining the shape function from the forcing parameters is applied here in ePBL as an example.  However, the function is not specific to ePBL and can also be used within other first order ocean surface boundary layer parameterizations \cite[such as KPP,][]{Large1994,Van2018kpp}. 

It is also important to consider that the neural network based model used in this work is not the only approach to find a relationship between the forcing terms and the vertical mixing profile.  The neural network is able to establish the existence of a relationship between its input and outputs, which is learned during the training process.  While the neural network can be applied in ocean models as-is to improve simulations, we also desire an in-depth understanding of the patterns in the inputs that the network used to make its skillful predictions.  In future work, we seek to relate the network's findings to the processes that govern the ocean surface boundary layer's behavior (e.g., with equation discovery).  This may ultimately lead to a simpler, interpretable and computationally low-cost physics based model for the shape function that can be learned from the neural network and applied in ocean models. 

\subsection{Implications for augmenting ocean parameterizations with Machine Learning}

A second implication of this work is demonstrating the potential for neural networks to improve parameterizations in ocean models. This implication is in agreement with several similar previous studies in earth system modeling \cite[e.g.][]{Gorman2018,yuval2020}.
As neural networks are not limited to individual processes, future avenues of research on ocean parameterizations will benefit from their usage.  For example, neural networks can be applied to incorporate different mixed layer processes such as non-local fluxes during convection, entrainement, Langmuir turbulence, symmetric instability,  surface wave effects, etc. into a single neural network model. Further improvements can be made which incorporate time history to improve predictions under transient forcings. Many existing ocean/atmosphere parameterizations have a physics based parent scheme with a few ad-hoc components or approximations. These components can be replaced or re-tuned using our approach or other emerging approaches such as Ensemble Kalman methods, posteriori criteria matching, etc. \cite[e.g.,][and references therein]{Gomez2022, Frezat2022}. Parameterizations in the form of weights and biases are advantageous because they can be re-tuned and further optimized to train as additional data, observations, and processes are presented.

The successful application of neural networks in an OGCM simulation unlocks the potential to test the importance of improving a certain process/parameterization in the model. For example, consider a case where the process studies' data exist, but a physics-based parameterization might be challenging to develop. Neural networks can parameterize that process and its impacts in an OGCM can be explored before going into a detailed parameterization development, which can be resource-consuming.   

One of the major sources of uncertainty in climate models arise from parameterizations due to their inadequate representation of sub-grid physics. Perhaps, high resolution or shorter time-steps can attenuate the effects of structural uncertainties in sub-grid parameterizations. Computational limitations often impose constraints on factors such as resolution, ensemble size, and integration time scales within models. These limitations underscore the need for improving the current generation of climate models, while steering away from relying on higher resolution models or shorter time steps. Combining traditional process-oriented studies with the emerging field of machine learning offers the potential for synergistic advancements, leading to the refinement of sub-grid models. We have established a pipeline whereby an existing parameterization is augmented to harness the capabilities of neural networks. The successful integration of neural network within the energetic Planetary Boundary Layer (ePBL), and its application to an ocean model, introduces opportunities for enhancing parameterizations that govern upper ocean mixing in climate models.

\appendix

\section{Why does $v_0$ change due to Coriolis parameter, $f$?}
\label{sec:appendixv0}

\begin{figure}[ht]
\includegraphics[width=\textwidth]{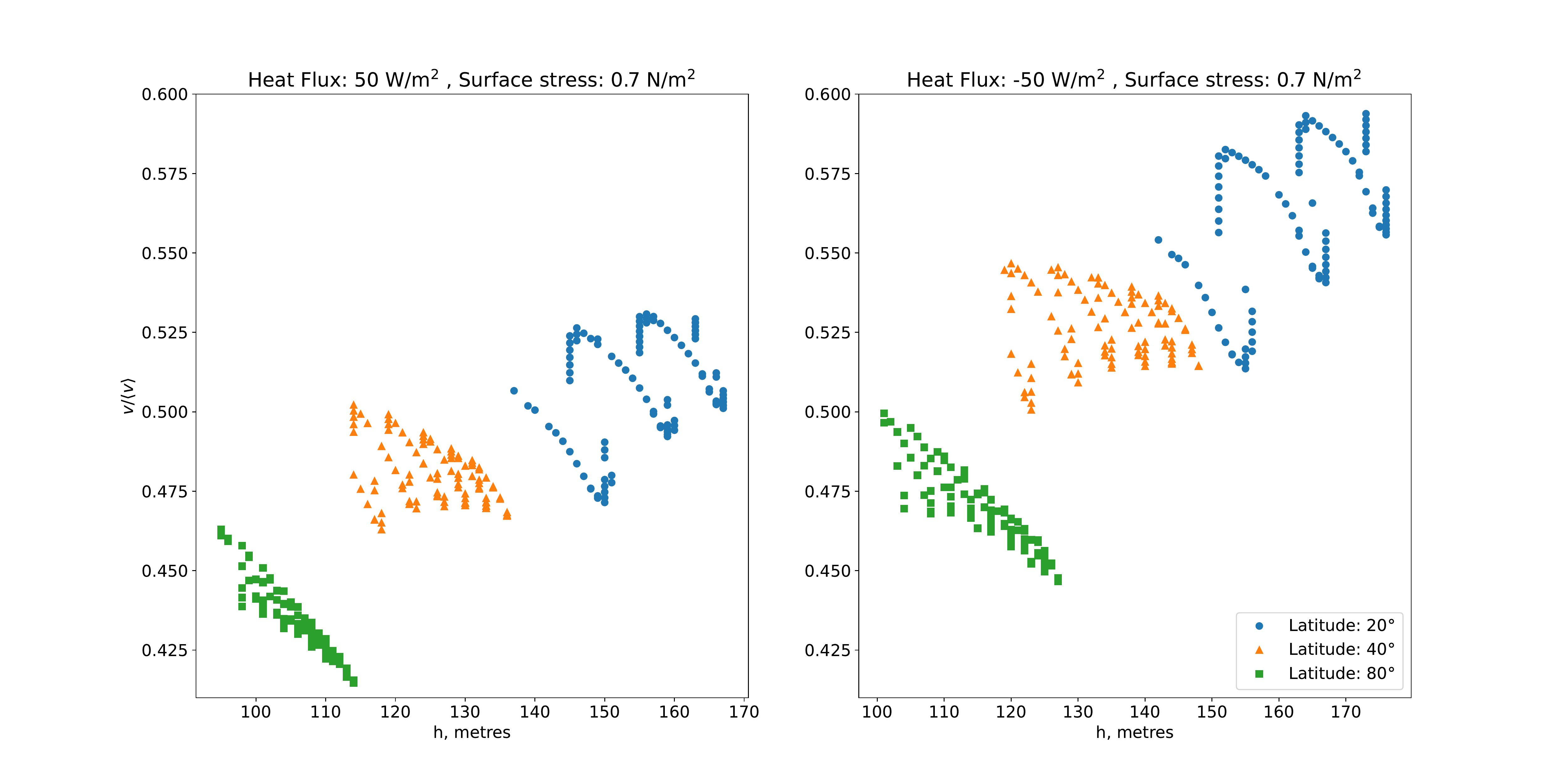}
\caption{Variation of normalized $v$ with repect to its mean value. $v_0$ varies due to heat flux, surface stress, latitude. Variations due to $h$ are within 5\% of the mean value of $v_0$ and hence it is reasonable to exclude $h$ from being an input to $\mathcal{N}_2$. Note that $v_0$ is a diagnosed quantity from the output of $k-\epsilon$ solely used to reconstruct the difusivity profile.}
\label{fig:variation_in_v}
\end{figure}

In general, turbulent velocity scales are related to turbulent kinetic energy and depend on boundary forcing, $u_*$ and $B_0$.  Here, in addition to $u_*$ and $B_0$, we find a dependency of the bulk turbulent velocity scale $v_0$ on $f$. The bulk velocity is diagnosed by using diffusivity and boundary layer depth from the training data as per Equation \ref{eq:v}. To predict $v_0$ using $\mathcal{N}_2$ the Coriolis parameter $f$ has been used because we found the model improves in ability to predict variations in $v_0$ in the training data. This is evident from Figure \ref{fig:variation_in_v}. 

Figure \ref{fig:variation_in_v} shows the variation of $v_0$ with respect to latitudes and $h$ under surface heating and cooling conditions. This indicates that $f$ is a useful input for accurately predicting $v_0$. Physically, the inclusion of $f$ is related to the role of rotation in limiting the wind-input of energy and the shear production of turbulence in the boundary layer through Ekman effects.  %
The variation due to $h$ is smaller than due to $f$, around 5\% of the mean value of $v_0$ for any particular set of forcing ($f, B_0, u_*$).  Since the implementation and generalization is significantly easier if the network only depends on external forcing parameters, we choose to include $f$ as an input to the network and neglect $h$.

\section{Quantifying uncertainty range covered in the forcing data}
\label{sec:appendixhx}
Table \ref{table:table1} gives the range of the forcing parameters covered in the training data set. A natural question is how much of the variability  observed in GCM simulations is covered in the training data.
We can estimate this using Shannon entropy \citep{shannon1948} which measures the amount of uncertainty and variability in a variable \citep{Sane2021, Sane2022, carcassi2021variability}. 

Shannon entropy of an event $x_i$ is given by $H(x) = \Sigma_{i=1}^{N} p_i log_{2} (1/p_i)$ \citep{cover1999} and measures the average amount of \textit{information} or surprise related to the event. We only use discrete probability distributions. Low probability events have high Shannon entropy because they cause more surprise compared to high probability events. 
It is a non-parametric measure and does not make any assumption about the distribution. $u_*$ and $B_0$ are non-Gaussian (Figure \ref{fig:variation_in_v}). 

For $u_*$: H($u_*$ $>$0.03 m/s) $\approx$ 95.5\%  and for $B_0$: H($|B_0|$ $>$ 2.1 $\times$ $10^{-7}$ m$^2$/s$^3$) $\approx$ 86\%. This can be interpreted as the values $u_*>0.03$ have 95.5\% uncertainty associated with them. So leaving out values of $u_*$ for which $u_*>0.03$ removes 95.5\% uncertainty from the training data.  
This is a simplistic estimate and assumes $u_*$ and $B_0$ are independent. These estimates show that our training data cover 95.5\% variability for $u_*$  and 86\% of $B_0$ as observed under realistic conditions in a GCM.

The training data points are uniform and although they cover most of the range seen in realistic conditions, the training data does not follow the same marginal probability distribution of $u_*$ and $B_0$ as well as the joint probability distribution between $u_*, B_0$. For machine learning application of parameterization development the consequence of sampling from joint distribution of variables from realistic conditions versus having uniformly spaced forcing is unknown as of now and will be left for future study.

\begin{figure}[ht!]
\includegraphics[width=\textwidth]{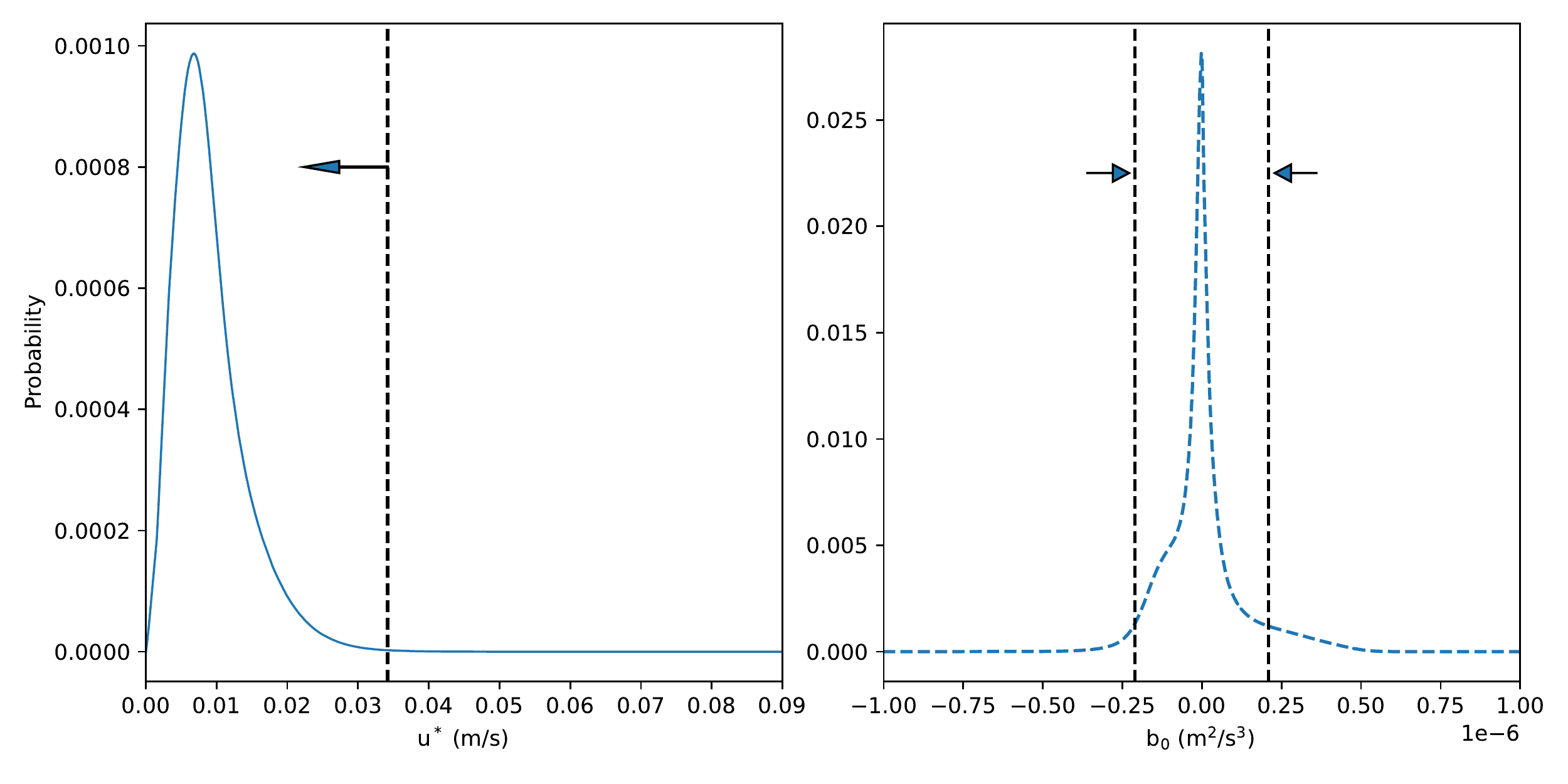}
\caption{Left: Probability density curve of surface friction velocity $u_*$. Right: Probability density curve of surface buoyancy flux $B_o$. The arrows denote the range covered in the training dataset.}
\end{figure}

\clearpage

\section{List of symbols and abbreviations}
\label{sec:appendixSym}

\begin{table}[h]
\centering
\caption{List of symbols}
\begin{tabular}{l  l  l }  %
\hline
Symbol & Decription  & Units (if applicable) \\ \hline
$\Psi$     & Generic output  & - \\ \hline
$\mathcal{F}$  & Generic Function & - \\ \hline
$\mathcal{N}$ & Neural Network function     & -  \\ \hline
$\bf{w}$ & hyperparameters in a Neural Network     & -  \\ \hline
$f$ & Coriolis parameter  & s$^{-1}$  \\ \hline
$w$ & vertical velocity & m/s \\ \hline
$u_*$ & Surface friction velocity  & ms$^{-1}$  \\ \hline
$b$ & Buoyancy  & ms$^{-2}$  \\ \hline
$B_0$ & Surface Buoyancy Flux $B_0=\overline{w^\prime b^\prime}_0$  & m$^2$s$^{-3}$  \\ \hline
$h$ & Boundary layer depth  & m  \\ \hline
$\phi$ & Generic tracer & - \\ \hline
$\kappa_{\phi}$ & Diffusivity of a variable $\phi$ & m$^2$ s$^{-1}$ \\ \hline
$b$ & Buoyancy  & ms$^{-2}$  \\ \hline
$L$ & Length scale used in diffusivity & m \\ \hline
$z$ & z co-ordinate, aligned with the local gravitational acceleration & m \\ \hline
$\sigma$ & sigma co-ordinate, defined by $z/h$ & - \\ \hline
$g(\sigma)$ & shape function which sets variation of diffusivity & - \\ \hline 
$v_0$ & velocity diagnosed from $k-\epsilon$ single column model runs & ms$^{-1}$ \\ \hline
MLD & Mixed layer depth & m \\ \hline
$k$ & turbulent kinetic energy & m$^2$/s$^2$ \\ \hline
$\epsilon$ & Dissipation of turbulent kinetic energy & m$^2$/s$^3$ \\ \hline
\end{tabular}
\label{table:table2}
\end{table}

\section*{Open Research Section}

The code and the data can be obtained from 
https://doi.org/10.5281/zenodo.8293998
The code includes scripts for generating the training data, training the neural network model, and code for vertical mixing scheme which has been modified to use neural networks. We have also provided code and data for plotting.

\acknowledgments

AS, AA and LZ received M$^2$LInES research funding through the generosity of Eric and Wendy Schmidt by recommendation of the Schmidt Futures program. AA was also supported by award NA18OAR4320123, from the National Oceanic and Atmospheric Administration (NOAA), U.S. Department of Commerce. The statements, findings, conclusions, and recommendations are those of the author(s) and do not necessarily reflect the views of the National Oceanic and Atmospheric Administration, or the U.S. Department of Commerce. We were intellectually supported by various other members of the M$^2$LInES project. We used the Stellar computational resources provided by Princeton University and the National Oceanic and Atmospheric Administration (NOAA) Geophysical Fluid Dynamics Laboratory (GFDL). We thank Dr. Enrico Zorzetto and Dr. Robert Hallberg for providing feedback for this article. We also thank Dr. Jun-Hong Liang and two anonymous reviewers for reviewing and providing profound insights, which led us to improve this manuscript. The authors thank the international Argo project and the various associated national programs for collecting and freely distributing the dataset. AS thanks his wife for showing remarkable grit in helping to improve the plain language summary.

\bibliography{refs}

\end{document}